\documentclass[sigconf]{acmart}

\AtBeginDocument{%
  }

\usepackage{tcolorbox}

\usepackage{bbding}
\usepackage{subfigure}
\usepackage[boxed,ruled,lined]{algorithm2e}
\usepackage{multirow}
\usepackage{makecell}

\newcommand{\argmin}{\operatornamewithlimits{arg\,min}}
\usepackage{arydshln} 
\usepackage{enumerate}
\usepackage{enumitem}
\usepackage{makecell}

\newcommand{\paratitle}[1]{\vspace{0.8ex}\noindent\textbf{#1}}

\newcommand{\srcandrec}{S\&R\xspace}
\newcommand{\ourname}{GSERec\xspace}

\setcopyright{acmlicensed}
\copyrightyear{2018}
\acmYear{2018}
\acmDOI{XXXXXXX.XXXXXXX}
\acmConference[Conference acronym 'XX]{Make sure to enter the correct
  conference title from your rights confirmation email}{June 03--05,
  2018}{Woodstock, NY}
\acmISBN{978-1-4503-XXXX-X/2018/06}

\begin{document}

\title{Benefit from Rich: Tackling Search Interaction Sparsity in \\ Search Enhanced Recommendation}

\author{Teng Shi}
\affiliation{
\institution{\mbox{Gaoling School of Artificial Intelligence}\\Renmin University of China}
  \city{Beijing}
  \country{China}
}
\email{shiteng@ruc.edu.cn}

\author{Weijie Yu}
\affiliation{
\institution{School of Information Technology and Management\\University of International Business and Economics}
  \city{Beijing}
  \country{China}
}
\email{yu@uibe.edu.cn}

\author{Xiao Zhang}
\affiliation{
\institution{\mbox{Gaoling School of Artificial Intelligence}\\Renmin University of China}
  \city{Beijing}
  \country{China}
}
\email{zhangx89@ruc.edu.cn}

\author{Ming He}
\affiliation{
\institution{AI Lab at Lenovo Research}
  \city{Beijing}
  \country{China}
}
\email{heming01@foxmail.com}

\author{Jianping Fan}
\affiliation{
\institution{AI Lab at Lenovo Research}
  \city{Beijing}
  \country{China}
}
\email{jfan1@lenovo.com}

\author{Jun Xu}
\affiliation{
\institution{\mbox{Gaoling School of Artificial Intelligence}\\Renmin University of China}
  \city{Beijing}
  \country{China}
}
\email{junxu@ruc.edu.cn}

\renewcommand{\shortauthors}{Teng Shi et al.}

\begin{abstract}
In modern online platforms, search and recommendation (\textbf{\srcandrec}) often coexist, offering opportunities for performance improvement through search-enhanced approaches. Existing studies show that incorporating search signals boosts recommendation performance.
However, the effectiveness of these methods relies heavily on rich search interactions. They primarily benefit a small subset of users with abundant search behavior, while offering limited improvements for the majority of users who exhibit only sparse search activity.
To address the problem of sparse search data in search-enhanced recommendation, we face two key 
challenges
: (1) how to learn useful search features for users with sparse search interactions,
and (2) how to design effective training objectives under sparse conditions.
Our idea is to leverage the features of users with rich search interactions to enhance those of users with sparse search interactions.
Based on this idea, we propose \textbf{\ourname}, a method that utilizes message passing on the User-Code \textbf{G}raphs to alleviate data sparsity in \textbf{S}earch-\textbf{E}nhanced \textbf{Rec}ommendation.
Specifically, we utilize Large Language Models (LLMs) with vector quantization to generate discrete codes, which connect similar users and thereby construct the graph.
Through message passing on this graph, embeddings of users with rich search data are propagated to enhance the embeddings of users with sparse interactions. To further ensure that the message passing captures meaningful information from truly similar users, we introduce a contrastive loss to better model user similarities. The enhanced user representations are then integrated into downstream search-enhanced recommendation models.
Experiments on three real-world datasets show that \ourname consistently outperforms baselines, especially for users with sparse search behaviors.

\end{abstract}



\maketitle

\begin{figure}[t]
    \centering
    \includegraphics[width=0.9\columnwidth]{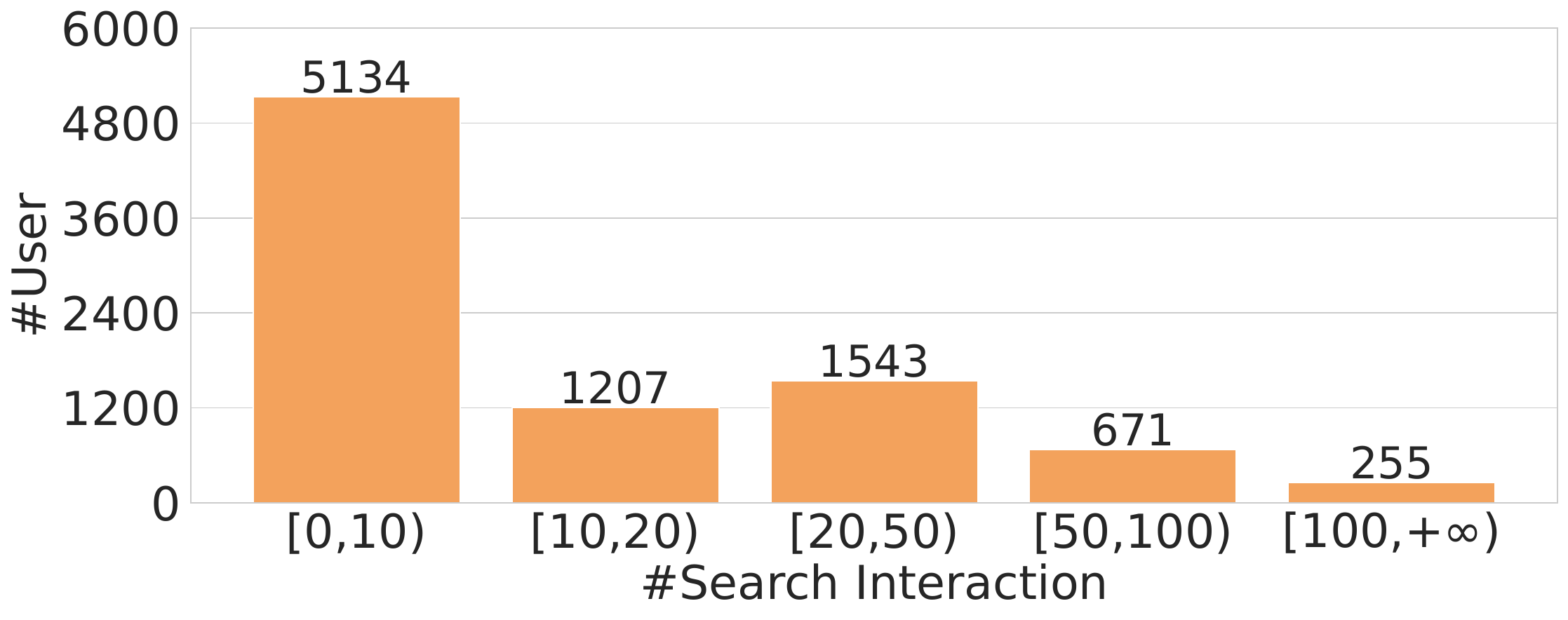}
   \vspace{-12px}
    \caption{User count statistics across different groups on the Qilin~\cite{chen2025qilin} dataset, where users are grouped by the number of their search interactions. 
    We observe that users with rich search interactions are few, while the majority are users with sparse interactions.}
    \label{fig:intro_sample_num}
    \vspace{-0.5cm}
\end{figure}

\section{Introduction}
Nowadays, many commercial apps offer both search and recommendation (\textbf{\srcandrec}) services to meet diverse user needs, such as e-commerce platforms (e.g., Taobao) and short video platforms (e.g., TikTok). 
In these scenarios, user \srcandrec behaviors frequently influence each other, providing an opportunity to enhance recommendation through search, allowing us to better model users with search behavior.

Existing search-enhanced recommendation methods primarily enhance the model from two perspectives:
(1)~Feature-level enhancement: 
These methods introduce additional search-related features on top of traditional recommendation models. For example, many approaches incorporate users’ search history into the model input~\cite{NRHUB,Query_SeqRec,IV4REC,SESRec}.
(2)~Loss-level enhancement: 
Many models~\cite{JSR,JSR2,UniSAR} adopt joint training of \srcandrec by introducing combined loss functions to simultaneously optimize both objectives, aiming to learn better user and item representations.

\begin{figure}[t]
    \centering
    \includegraphics[width=0.98\columnwidth]{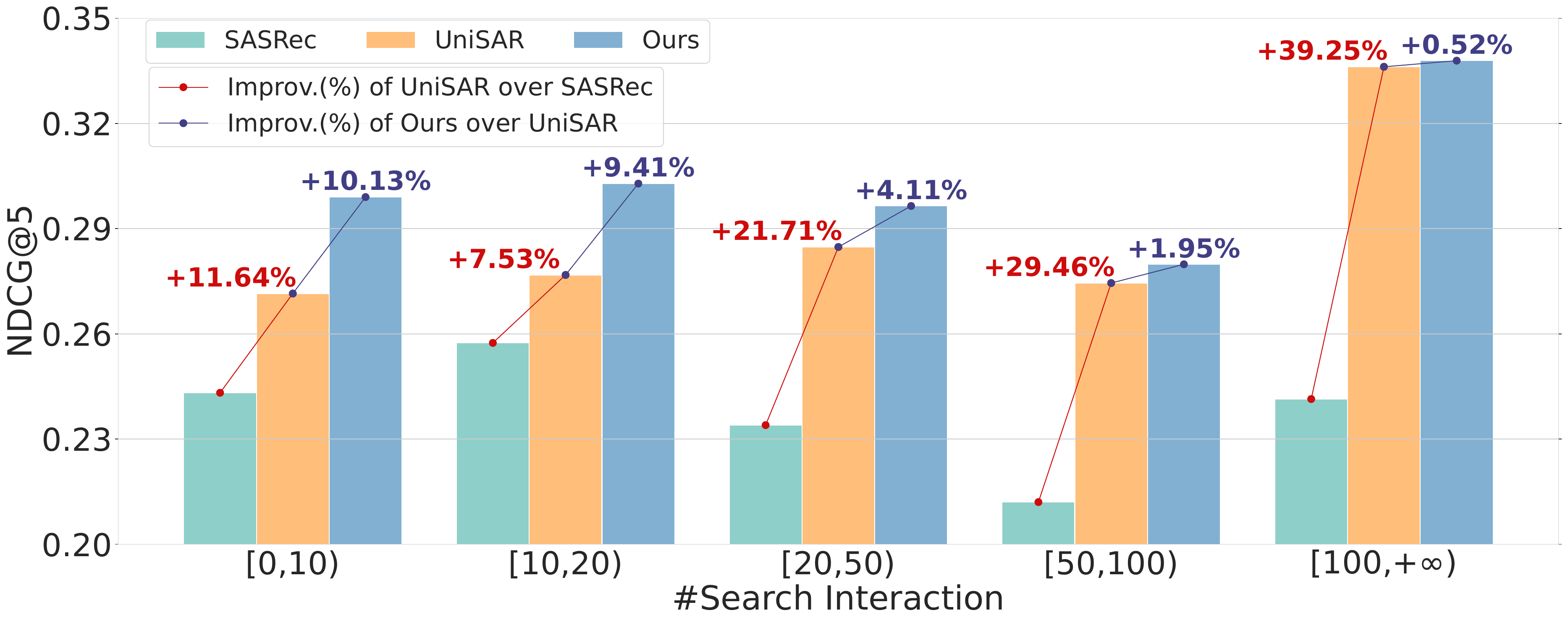}
   \vspace{-12px}
    \caption{
    Relative improvements of the state-of-the-art model UniSAR~\cite{UniSAR} over the traditional recommendation model SASRec~\cite{SASREC} across different user groups on the Qilin dataset, along with the improvements of our model over UniSAR. User groups are defined using the same strategy as in Figure~\ref{fig:intro_sample_num}. UniSAR shows greater improvements for users with more search interactions, while our model effectively alleviates data sparsity and achieves larger gains for users with fewer search interactions.
    }
    \label{fig:intro_improve}
    \vspace{-0.5cm}
\end{figure}

While these methods have achieved promising results, their 
enhancement remains constrained by data sparsity. For example, in search-enhanced recommendation, users with limited search interactions contribute minimally in two ways: (1) the search history features incorporated into the model are scarce; and (2) the loss derived from their sparse search data has limited effect on optimizing representations during joint training. 
As a result, existing models yield greater improvements for users with rich search interactions, while offering limited benefits for users with sparse search behavior.
However, as shown in Figure~\ref{fig:intro_sample_num}, users with rich search interactions are few, while the majority of users exhibit sparse search behavior.
Figure~\ref{fig:intro_improve} further reveals that the state-of-the-art search-enhanced recommendation model UniSAR~\cite{UniSAR} achieves greater improvements primarily for users with richer search interactions.

Alleviating the issue of sparse user search data in search-enhanced recommendation presents two main challenges:
(1)~How to effectively enhance features for users with sparse data. 
For example, in search-enhanced recommendation, 
how to derive informative search features for users with limited search interactions;
(2)~How to design improved loss functions during training to ensure that the representations of users with sparse data can still be well optimized.
Our key idea is that not all users suffer from search sparsity—some have rich search interactions. We enhance the features of users with sparse search behaviors by propagating information from similar users with rich search histories.

Based on this idea, to address the above issues, we propose a method called \textbf{\ourname}, which performs message passing on the User-Code \textbf{G}raphs to enhance the representations of users with sparse search interactions, thereby alleviating data sparsity in \textbf{S}earch-\textbf{E}nhanced \textbf{Rec}ommendation.
Specifically, we first utilize a Large Language Model (LLM)~\cite{zhao2023survey} to summarize users’ \srcandrec preferences. These preferences are then encoded using an embedding model and transformed into discrete codes via vector quantization~\cite{zeghidour2021soundstream,TIGER}. Next, we connect each user to their corresponding codes, and the shared codes link similar users together, forming the graph.
Subsequently, by performing message passing on this graph, the embeddings of users with rich search interactions can be propagated to enhance the embeddings of users with sparse search interactions.
Furthermore, to ensure that message passing captures useful information from similar users, we design contrastive learning~\cite{chen2020simple,he2020momentum} objectives to help user embeddings better capture the similarity between users.
Finally, the enhanced user representations are integrated with the \srcandrec history features in downstream search-enhanced recommendation models for the final recommendation~task.

The major contributions of the paper are summarized as follows:

\noindent\textbf{$\bullet $}~We identify a key limitation of existing search-enhanced recommendation methods: their performance is limited for users with sparse search histories. This highlights the challenge of extracting informative search representations and designing effective loss functions under data sparsity.

\noindent\textbf{$\bullet $}~We propose \ourname, which performs message passing on the User-Code graphs to enhance the embeddings of users with sparse search interactions, thereby alleviating the data sparsity problem in search-enhanced recommendation. Furthermore, we design contrastive learning objectives to better model user similarity, thereby enabling the message passing process to extract more informative signals.

\noindent\textbf{$\bullet $}~Experimental results on three datasets validate the effectiveness of \ourname: it not only outperforms traditional recommendation methods but also surpasses existing search-enhanced recommendation approaches. Moreover, as shown in Figure~\ref{fig:intro_improve}, \ourname achieves notably larger gains for users with sparse search interactions.
\section{Related Work}

\paratitle{Recommendation.}
Recommender systems~\cite{zhang2024model,zhang2024modeling,zhang2024qagcf,zhang2025test,zhang2024reinforcing,dai2024modeling,shen2024generating} suggest items aligned with users' interests by modeling their preferences. Sequential recommendation~\cite{GRU4REC, DIN, DIEN, FMLPREC, LRURec, SAQRec,tang2025think} focuses on capturing user interests based on historical interactions. SASRec~\cite{SASREC} and BERT4Rec~\cite{BERT4REC} use Transformer architectures~\cite{vaswani2017attention} to model sequential behaviors, while CL4SRec~\cite{xie2022contrastive} incorporates contrastive learning~\cite{chen2020simple, he2020momentum} to enhance user history representations.
Graph-based approaches~\cite{wu2021self, yu2022graph,lin2022improving,lightgcl,tang2024towards} model users through collaborative relationships between users and items. LightGCN~\cite{he2020lightgcn} aggregates user-item interactions via neighborhood information, while SGL~\cite{wu2021self} and SimGCL~\cite{yu2022graph} leverage contrastive learning to improve GNN training.
Recently, the integration of large language models (LLMs) into recommender systems has gained attention~\cite{liu2024llm, ren2024representation,liu2024discrete,dai2023uncovering,qin2024enhancing}. KAR~\cite{xi2024towards} enhances recommendations using precomputed vectors of user reasoning and item knowledge from LLMs, and LLM-ESR~\cite{liu2024llm} improves sequential recommendation by combining LLM-generated semantics with collaborative signals to address long-tail challenges.
In contrast to these methods, this paper proposes improving recommender systems by leveraging search data to enhance model performance.

\begin{figure*}[t]
    \centering
        \includegraphics[width=0.95\linewidth]{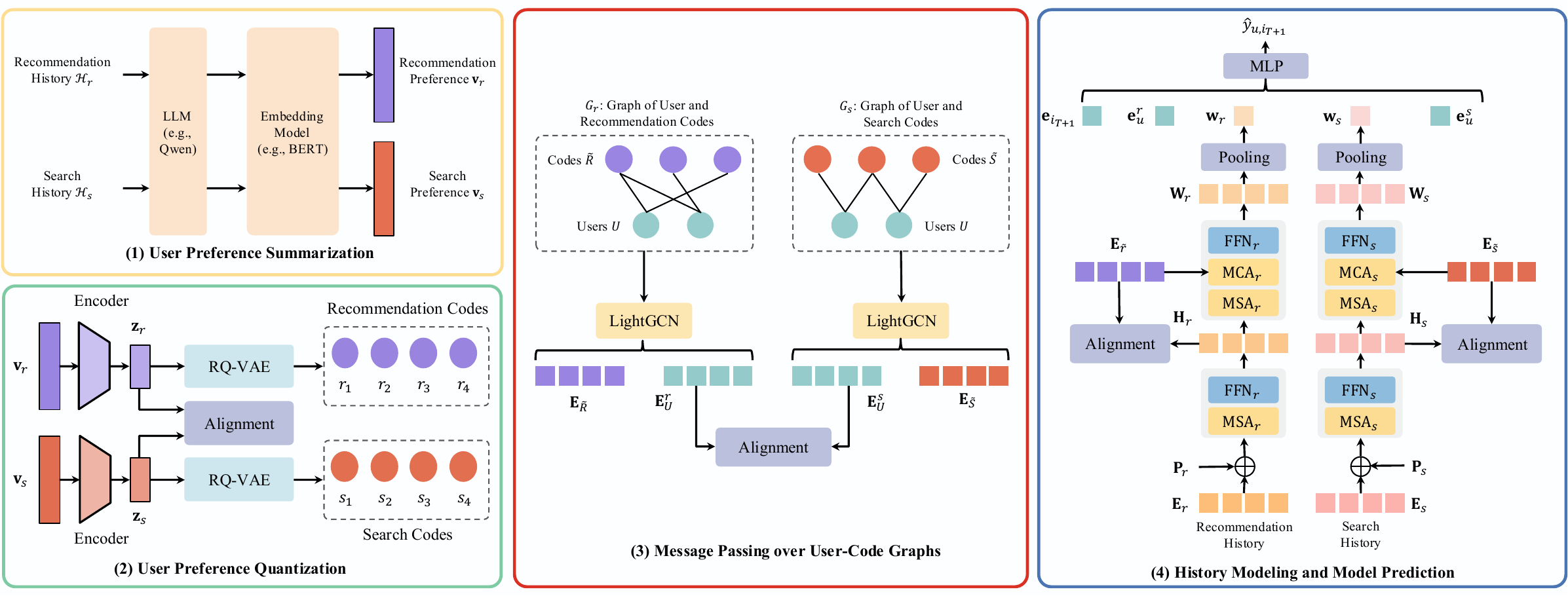}
   \vspace{-12px}
    \caption{
    The overall framework of \ourname.
    The framework consists of two stages:
    User-Code Graph Construction:
    (1)~User Preference Summarization;
    (2)~User Preference Quantization.
    Search Enhanced Recommendation Modeling:
    (3)~Message Passing over User-Code Graph;
    (4)~History Modeling and Prediction.
    }
\label{fig:method}
\vspace{-0.5cm}
\end{figure*}

\paratitle{Search Enhanced Recommendation.}
In recent years, the use of search~\cite{zhang2025trigger3,qin2025similarity,shi2025retrieval,shen2024survey,qin2024explicitly,qin2025maps,qin2025uncertainty,zhang2025syler,sun2024logic,zhang2024citalaw}  data to enhance recommendation model performance has garnered significant attention~\cite{IV4REC, wang2024enhancing, JSR, SRJgraph, xie2024unifiedssr, zhang2024unified, UniSAR, shi2025unified,zhao2025unifying,penha2024bridging}. JSR~\cite{JSR, JSR2} jointly trains two models for \srcandrec using a shared loss function. USER~\cite{USER} integrates both user \srcandrec behaviors, which are then processed through a transformer encoder. SESRec~\cite{SESRec} employs contrastive learning to distinguish between similar and dissimilar interests in user \srcandrec behaviors. UnifiedSSR~\cite{xie2024unifiedssr} introduces a dual-branch network to simultaneously encode product and query histories. UniSAR~\cite{UniSAR} models user transition behaviors between \srcandrec using transformers with distinct masking mechanisms and contrastive learning.
Unlike these methods, we alleviate the sparsity issue in search-enhanced recommendation by enriching sparse user embeddings through message~passing.

\section{Problem Formulation}
We denote the sets of users, items, and queries as $\mathcal{U}$, $\mathcal{I}$, and $\mathcal{Q}$, respectively. Each user $u \in \mathcal{U}$ has a chronologically ordered recommendation history $\mathcal{H}_{r}=\{i_1,i_2,\ldots,i_{N_r}\}$ and a search history $\mathcal{H}_{s}=\{(q_1,\mathcal{I}_{q_1}),(q_2,\mathcal{I}_{q_2}),\ldots,(q_{N_s},\mathcal{I}_{q_{N_s}})\}$,
where $N_r$ and $N_s$ denote the lengths of the user's recommendation and search histories, respectively. The total number of user interactions is denoted as $T=N_r+N_s$. Here, $i_k \in \mathcal{I}$ is the $k$-th item the user interacted with, $q_k \in \mathcal{Q}$ is the $k$-th query issued by the user, and $\mathcal{I}_{q_k} = \{i_1,i_2,\dots,i_{N_{q_k}}\}$ is the set of $N_{q_k}$ items clicked by the user after searching query $q_k$.
Our goal is to train a recommendation model $\Theta$ that predicts the next item $i_{T+1}$ based on the user's recommendation history $\mathcal{H}_r$ and search history $\mathcal{H}_{s}$.

Existing search-enhanced recommendation methods tend to yield greater improvements for users with rich search interactions, while offering limited benefits for those with sparse search behavior. To address this imbalance, we aim to enhance the representations of users with sparse search interactions by propagating information from users with richer search histories, thereby alleviating the data sparsity issue in search-enhanced recommendation.

\section{Our Approach}

This section introduces our method, \ourname, illustrated in Figure~\ref{fig:method}, which includes two main components: \textbf{User-Code Graph Construction} (\S~\ref{sec:user_code_graph}) and 
\textbf{Search Enhanced Recommendation Modeling}
(\S~\ref{sec:sequential_model}).
\textbf{User-Code Graph Construction} includes:
(1)~User Preference Summarization (\S~\ref{sec:user_prefer_summarize}): summarizes users’ \srcandrec preferences using the LLM;
(2)~User Preference Quantization (\S~\ref{sec:user_quant}): discretizes the summarized preferences into codes via vector quantization;
(3)~Graph Construction (\S~\ref{sec:graph_construct}): connects similar users through shared codes to form the user-code bipartite graphs.
\textbf{Search Enhanced Recommendation Modeling} includes:
(1)~Message Passing over the User-Code Graph (\S~\ref{sec:message_passing}): enhances the representations of users with sparse search interactions by propagating information from users with richer interactions;
(2)~Historical Modeling (\S~\ref{sec:history_model}): integrates the enhanced user representations with their \srcandrec histories for final prediction.

\subsection{User-Code Graph Construction}
\label{sec:user_code_graph}
This section introduces the construction of the user-code graph. We first use the LLM to summarize users’ \srcandrec preferences. These preferences are then encoded using an embedding model and discretized into codes via vector quantization. Finally, each user is connected to their corresponding codes, and users who share similar codes are linked, forming the user-code graph.

\subsubsection{User Preference Summarization}
\label{sec:user_prefer_summarize}
For each user $u \in \mathcal{U}$, we input her search history $\mathcal{H}_s$ and recommendation history $\mathcal{H}_r$ into a LLM to summarize her \srcandrec preferences. The prompts provided to the LLM are as follows:
\begin{center}
\begin{tcolorbox}[colframe=black!75!white, title=Search Preference Summarization,width=0.95\columnwidth]
\footnotesize
\textbf{Prompt:} 
Please analyze the queries and clicked items in the user's search history, and summarize the user's interest topics, areas of focus, style tendencies, or preference types. 
Here is the user's search history \emph{\{history\}}, where each record contains the user's query and the items the user clicked on under that query.
\end{tcolorbox}
\end{center}

\begin{center}
\begin{tcolorbox}[colframe=black!75!white, title=Recommendation Preference Summarization,width=0.95\columnwidth]
\footnotesize
\textbf{Prompt:} 
Please analyze the provided user recommendation history and summarize the user's possible interests, style tendencies, and preferred item types.
Here is the user's recommendation history \emph{\{history\}}, where each record represents an item the user has clicked on.
\end{tcolorbox}
\end{center}
The user’s \srcandrec preferences, as summarized by the LLM, are individually encoded using a pretrained embedding model (e.g., BERT~\cite{devlin-etal-2019-bert} or BGE~\cite{xiao2024c,bge-m3}) to obtain dense representations, 
denoted as $\mathbf{v}_s \in \mathbb{R}^{d_e}$ and $\mathbf{v}_r \in \mathbb{R}^{d_e}$. 
Here, $d_e$ represents the dimensionality of the embedding model.
It is important to note that the embedding model is pretrained and remains frozen during the entire training~process.

\subsubsection{User Preference Quantization}
\label{sec:user_quant}
We discretize the encoded user preferences into codes to facilitate subsequent graph construction. Specifically, we adopt Residual Quantized Variational Autoencoder (RQ-VAE)~\cite{zeghidour2021soundstream,TIGER,LC_Rec}, a widely used vector quantization method.
The user's \srcandrec preferences are first encoded using two separate~encoders:
\begin{equation*}
    \mathbf{z}_s = \mathrm{Encoder}_s(\mathbf{v}_s), \quad
    \mathbf{z}_r = \mathrm{Encoder}_r(\mathbf{v}_r), 
\end{equation*}
where $\mathbf{z}_s,\mathbf{z}_r \in \mathbb{R}^{d_l}$ 
denote the latent representations of the user's \srcandrec preferences, respectively, and $d_l$ is the dimension of the embedding space. $\mathrm{Encoder}_s(\cdot)$ and $\mathrm{Encoder}_r(\cdot)$ are implemented as multilayer perceptrons (MLPs).

$\mathbf{z}_s$ and $\mathbf{z}_r$ 
encode the \srcandrec preferences of the same user. To better model the similarity between different users, we align them before quantization.
To this end, we apply contrastive learning by minimizing the following InfoNCE~\cite{oord2018representation} loss:
\begin{equation}
\label{eq:rq_user_align}
\begin{aligned}
    \mathcal{L}_{\mathrm{RQ\text{-}CL}}=
    - & \left[ \mathrm{log}\frac{\mathrm{exp}(\mathrm{sim}(\mathbf{z}_s,\mathbf{z}_r)/\tau_1)}{\sum_{z_{r}^{-} \in \mathcal{Z}_{r}^{\mathrm{neg}}}\mathrm{exp}(\mathrm{sim}(\mathbf{z}_s,\mathbf{z}_{r}^{-})/\tau_1)} \right.\\
    & \left. +~~\mathrm{log}\frac{\mathrm{exp}(\mathrm{sim}(\mathbf{z}_s,\mathbf{z}_r)/\tau_1)}{\sum_{z_{s}^{-} \in \mathcal{Z}_{s}^\mathrm{neg}}\mathrm{exp}(\mathrm{sim}(\mathbf{z}_s^{-},\mathbf{z}_r)/\tau_1)} \right],
\end{aligned}
\end{equation}
where $\mathrm{sim(\cdot)}$ denotes a similarity function (e.g., cosine similarity), $\tau_1$ is a learnable temperature coefficient.
$\mathcal{Z}_{r}^{\mathrm{neg}}$ and $\mathcal{Z}_{s}^\mathrm{neg}$ denote the negative samples from other users within the same batch.

Next, 
$\mathbf{z}_s$ and $\mathbf{z}_r$ 
are independently discretized into $L$ codes using two separate $L$-level codebooks.
Taking the quantization of the user's search preferences as an example, at each level 
$l \in \{1,2,\ldots,L\}$, we define a codebook 
$\mathcal{CS}_l = \{\mathbf{e}_k\}_{k=1}^{N_c}$,
where $N_c$ is the size of each codebook and 
$\mathbf{e}_k \in \mathbb{R}^{d_l}$ 
is a learnable code embedding. The residual quantization process is as follows:
\begin{equation}
\label{eq:rq_process}
\begin{cases}
s_l = \argmin_{k} || \mathbf{r}_{l-1}^s - \mathbf{e}_{k}||^2_2, \quad \mathbf{e}_{k} \in \mathcal{CS}_l, \\
\mathbf{r}_{l}^s = \mathbf{r}_{l-1}^s - \mathbf{e}_{s_l}, \quad \mathbf{r}_0 = \mathbf{z}_s \in \mathbb{R}^{d_l}, \\
\end{cases}
\end{equation}
where $s_l$ denotes the index of the selected code at level $l$,
and $\mathbf{r}_{l-1}^s$ is the residual from the previous level.

Through the recursive quantization process described in Eq.~\eqref{eq:rq_process}, we obtain the discrete codes 
$\tilde{s}$ and the quantized embedding 
$\hat{\mathbf{z}}_s = \sum_{l=1}^{L}\mathbf{e}_{s_l}$ 
for the user's search preference.
Similarly, we can obtain the discrete codes 
$\tilde{r}$
and the quantized embedding 
$\hat{\mathbf{z}}_r = \sum_{l=1}^{L}\mathbf{e}_{r_l}$ 
for the user's recommendation preference.
The discrete codes $\tilde{s}$ and $\tilde{r}$ are as~follows:
\begin{equation}
\label{eq:user_code}
    \tilde{s}=\left[s_1,s_2,\ldots,s_{L}\right],\quad
    \tilde{r}=\left[r_1,r_2,\ldots,r_{L}\right]
\end{equation}
The quantized embeddings 
$\hat{\mathbf{z}}_s$ and $\hat{\mathbf{z}}_r$ 
are then passed through two separate decoders to reconstruct the original user \srcandrec preference representations, 
$\mathbf{v}_s$ and $\mathbf{v}_r$, 
respectively:
\begin{equation*}
    \hat{\mathbf{v}}_s = \mathrm{Decoder}_s(\hat{\mathbf{z}}_s), \quad
    \hat{\mathbf{v}}_r = \mathrm{Decoder}_r(\hat{\mathbf{z}}_r), 
\end{equation*}
where $\mathrm{Decoder}_s(\cdot)$ and $\mathrm{Decoder}_r(\cdot)$ denote two MLPs.
The reconstruction loss for training the encoders and decoders is calculated~as:
\begin{equation}
    \mathcal{L}_{\text{Recon}} = ||\mathbf{v}_s - \hat{\mathbf{v}}_s||^2_2 + ||\mathbf{v}_r - \hat{\mathbf{v}}_r||^2_2. \\
\end{equation}
To optimize the quantization process, we further introduce the residual quantization loss, which is formulated as:
\begin{equation}
\begin{cases}
\mathcal{L}_{\text{RQ}}^{s} = \sum_{l=1}^{L} ||\mathrm{sg}[\mathbf{r}_{l-1}^s] - \mathbf{e}_{s_l}||^2_2 + ||\mathbf{r}_{l-1}^s - \mathrm{sg}[\mathbf{e}_{s_l}] ||^2_2,  \\
\mathcal{L}_{\text{RQ}}^{r} = \sum_{l=1}^{L} ||\mathrm{sg}[\mathbf{r}_{l-1}^r] - \mathbf{e}_{r_l}||^2_2 + ||\mathbf{r}_{l-1}^r - \mathrm{sg}[\mathbf{e}_{r_l}] ||^2_2,  \\
\mathcal{L}_{\text{RQ}}= \mathcal{L}_{\text{RQ}}^{s} + \mathcal{L}_{\text{RQ}}^{r}, \\
\end{cases}
\end{equation}
where $\mathrm{sg}[\cdot]$ indicates the stop-gradient operation.
The loss $\mathcal{L}_{\text{RQ}}$ is employed to optimize the code embeddings across all codebooks. 
Finally, the total objective for user preference quantization combines the reconstruction loss, quantization loss, and contrastive loss~as:
\begin{equation}
\label{eq:rq_total_loss}
\mathcal{L}_{\text{RQ-VAE}} =\mathcal{L}_{\text{Recon}} + \lambda_{\mathrm{RQ}} \mathcal{L}_{\text{RQ}} + \lambda_{\mathrm{RQ\text{-}CL}} \mathcal{L}_{\mathrm{RQ\text{-}CL}},
\end{equation}
where $\lambda_{\mathrm{RQ}}$ and $\lambda_{\mathrm{RQ-CL}}$ are hyper-parameters controlling the contributions of the respective loss components.

\begin{figure}[t]
    \centering
        \includegraphics[width=0.75\columnwidth]{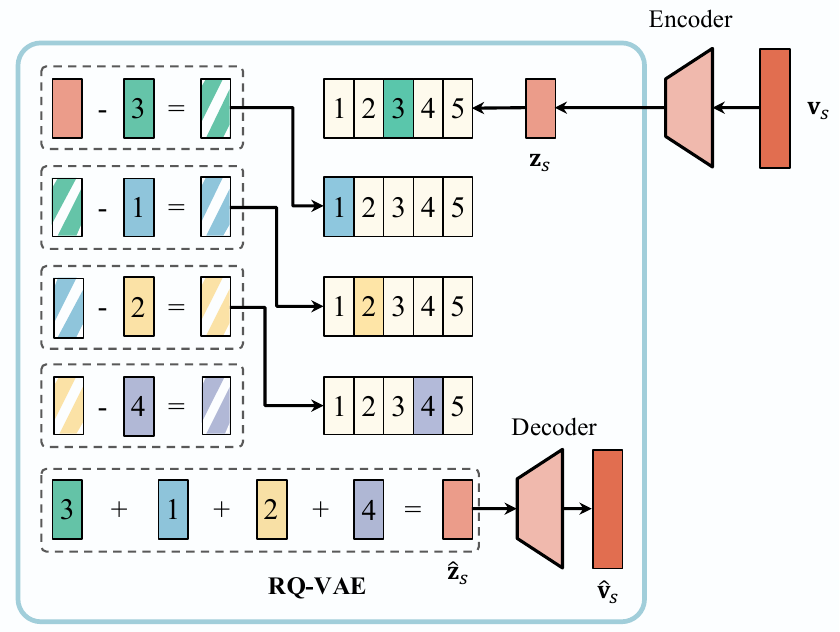}
   \vspace{-8px}
    \caption{
    The Residual Quantized Variational Autoencoder (RQ-VAE) process. We illustrate the procedure using the quantization of the search preference embedding $\mathbf{v}_s$ as an example.
    }
\label{fig:rqvae}
\vspace{-0.5cm}
\end{figure}

\subsubsection{Graph Construction}
\label{sec:graph_construct}
After quantizing user \srcandrec preferences into discrete codes, we construct two bipartite graphs to model the relationships between users and their corresponding \srcandrec code representations. Specifically, let 
$\widetilde{\mathcal{S}}$ and $\widetilde{\mathcal{R}}$
denote the sets of search codes and recommendation codes, respectively. The affiliation matrices between users and these codes are defined as 
$\mathbf{AS} \in \{0,1\}^{|\mathcal{U}| \times |\widetilde{\mathcal{S}}|}$ and 
$\mathbf{AR} \in \{0,1\}^{|\mathcal{U}| \times |\widetilde{\mathcal{R}}|}$, 
where $\mathbf{AS}_{u,s}=1$ 
indicates that user $u$ is associated with the search code $s$, 
and similarly for 
$\mathbf{AR} $ 
with recommendation codes.

Based on the affiliation matrices, we construct two bipartite graphs: 
$\mathcal{G}_s=\{\mathcal{V}_s,\mathcal{E}_s\}$ for search preferences and 
$\mathcal{G}_r=\{\mathcal{V}_r,\mathcal{E}_r\}$ for recommendation preferences.
The node sets are defined as $\mathcal{V}_s = \mathcal{U} \cup \widetilde{\mathcal{S}}$ and $\mathcal{V}_r = \mathcal{U} \cup \widetilde{\mathcal{R}}$.
The edge sets are given by 
$\mathcal{E}_s=\{(u,s)|u\in \mathcal{U}, s\in \widetilde{\mathcal{S}}, \mathbf{AS}_{u,s}=1\}$ 
and 
$\mathcal{E}_r=\{(u,r)|u\in \mathcal{U}, r\in \widetilde{\mathcal{R}}, \mathbf{AR}_{u,r}=1\}$, 
where an edge indicates an affiliation between a user and a corresponding preference code.

The constructed bipartite graphs are utilized to enhance user embeddings for subsequent search enhanced recommendation modeling, which will be detailed in the next section.

\subsection{\mbox{Search Enhanced Recommendation Modeling}}

\label{sec:sequential_model}
This section introduces the 
Search Enhanced Recommendation Modeling
module. We first apply message passing over the user-code graph to enhance the embeddings of users with sparse search interactions by leveraging information from users with rich search interactions. Then, the enhanced user and code embeddings are integrated with the \srcandrec histories in the downstream search-enhanced recommendation model for final prediction.

\subsubsection{Embedding Layer}
We maintain three embedding tables to represent users, items, and query words: 
$\mathbf{E}_{\mathcal{U}} \in \mathbb{R}^{|\mathcal{U}| \times d}$, 
$\mathbf{E}_{\mathcal{I}} \in \mathbb{R}^{|\mathcal{I}| \times d}$, and 
$\mathbf{E}_{\mathcal{W}} \in \mathbb{R}^{|\mathcal{W}| \times d}$, respectively. 
Here, $\mathcal{W}$ denotes the vocabulary comprising all words appearing in user queries, and $d$ is the embedding dimension. 
Given a specific user $u$ and item $i$, their corresponding embeddings $\mathbf{e}_u \in \mathbb{R}^{d}$ and $\mathbf{e}_i \in \mathbb{R}^{d}$ are retrieved via standard lookup operations.
For a query $q$ composed of a sequence of words $\{ w_1, w_2, \dots, w_{|q|}\} \subseteq \mathcal{W}$, we follow prior work~\cite{SESRec,UniSAR} and represent the query by averaging its constituent word embeddings: $\mathbf{e}_q = \mathrm{Mean}(\mathbf{e}_{w_1},\mathbf{e}_{w_2}, \dots,\mathbf{e}_{w_{|q|}}) \in \mathbb{R}^{d}$, where $\mathbf{e}_{w_i} \in \mathbb{R}^{d}$ denotes the embedding of the $i$-th word in the query.

In addition, we introduce two embedding tables, 
$\mathbf{E}_{\widetilde{\mathcal{S}}}^{(0)} \in \mathbb{R}^{|\widetilde{\mathcal{S}}| \times d}$ and 
$\mathbf{E}_{\widetilde{\mathcal{R}}}^{(0)} \in \mathbb{R}^{|\widetilde{\mathcal{R}}| \times d}$, to represent the discrete codes corresponding to \srcandrec preferences, respectively.
The initial embedding of a search code $s \in \widetilde{\mathcal{S}}$ or a recommendation code $r \in \widetilde{\mathcal{R}}$, denoted as $\mathbf{e}_{s}^{(0)}$ and $\mathbf{e}_{r}^{(0)}$, is obtained via standard embedding lookup from the respective~tables.

\subsubsection{Message Passing over User-Code Graphs}
\label{sec:message_passing}
To leverage the representations of users with rich search interactions to enhance those of users with sparse interactions, we perform message passing on the user-code graph defined in \S~\ref{sec:graph_construct}.
In these graphs, users are connected via shared preference codes, allowing semantically similar users to exchange information and mutually enhance their representations. Specifically, we adopt LightGCN~\cite{he2020lightgcn} as the propagation framework, leveraging its simplified yet effective design to iteratively refine user and code embeddings through neighborhood~aggregation.

Taking the propagation over the graph $\mathcal{G}_s$ as an example, the embeddings are iteratively updated through $K$ layers of message passing. 
Let $\mathbf{e}_u^{S(k)}$ and $\mathbf{e}_{s}^{S(k)}$ denote the embeddings of user $u$ and search code $s \in \widetilde{\mathcal{S}}$ at the $k$-th layer, respectively, where the initial embeddings are given by $\mathbf{e}_u^{S(0)}=\mathbf{e}_u$ and $\mathbf{e}_{s}^{S(0)}=\mathbf{e}_{s}^{(0)}$.
The update rule at the $k$-th propagation layer is defined as:
\begin{small}
\begin{equation}
\label{eq:lightgcn}
\begin{aligned}
    \mathbf{e}_u^{S(k)} &= \sum_{s \in \mathcal{N}_u} \frac{1}{\sqrt{|\mathcal{N}_u||\mathcal{N}_{s}|}} \cdot \mathbf{e}_{s}^{S(k-1)}, \\
    \mathbf{e}_{s}^{S(k)} &= \sum_{u \in \mathcal{N}_{s}} \frac{1}{\sqrt{|\mathcal{N}_u||\mathcal{N}_{s}|}} \cdot \mathbf{e}_u^{S(k-1)},
\end{aligned}
\end{equation}
\end{small}
where $\mathcal{N}_u$ and $\mathcal{N}_{s}$ denote the neighboring codes of user $u$ and the neighboring users of code $s$, respectively. The final embeddings are obtained by aggregating embeddings from all layers:
\begin{small}
\begin{equation}
\label{eq:lightgcn_final}
\mathbf{e}_u^{s} = \frac{1}{K+1} \sum_{k=0}^{K} \mathbf{e}_u^{S(k)}, \quad
\mathbf{e}_{s} = \frac{1}{K+1} \sum_{k=0}^{K} \mathbf{e}_{s}^{S(k)},
\end{equation}
\end{small}
where $\mathbf{e}_u^s, \mathbf{e}_{s} \in \mathbb{R}^d$ are the final representations for user $u$ and code $s$, respectively.
In a similar manner, message passing is performed over the graph $\mathcal{G}_r$ to obtain the final representations $\mathbf{e}_u^{r}$ and $\mathbf{e}_{r}$ for user $u$ and recommendation code $r \in \widetilde{\mathcal{R}}$.
Then, we obtain the enhanced \srcandrec embeddings for all users:
\begin{small}
\begin{equation*}
\mathbf{E}_{\mathcal{U}}^s = [\mathbf{e}_{u_1}^s, \mathbf{e}_{u_2}^s, \ldots, \mathbf{e}_{u_{|\mathcal{U}|}}^s]^{\mathsf{T}},\quad
\mathbf{E}_{\mathcal{U}}^r = [\mathbf{e}_{u_1}^r, \mathbf{e}_{u_2}^r, \ldots, \mathbf{e}_{u_{|\mathcal{U}|}}^r]^{\mathsf{T}},
\end{equation*}
\end{small} 
where $\mathbf{E}_{\mathcal{U}}^s,\mathbf{E}_{\mathcal{U}}^r \in \mathbb{R}^{|\mathcal{U}| \times d}$ denote the user's \srcandrec embeddings, respectively.
Similarly, we obtain the code embeddings for \srcandrec:
\begin{small}
\begin{equation*}
\mathbf{E}_{\widetilde{\mathcal{S}}} = [\mathbf{e}_{s_1}, \mathbf{e}_{s_2}, \ldots, \mathbf{e}_{s_{|\widetilde{\mathcal{S}}|}}]^{\mathsf{T}},\quad
\mathbf{E}_{\widetilde{\mathcal{R}}} = [\mathbf{e}_{r_1}, \mathbf{e}_{r_2}, \ldots, \mathbf{e}_{r_{|\widetilde{\mathcal{R}}|}}]^{\mathsf{T}}, 
\end{equation*}
\end{small}
where $\mathbf{E}_{\widetilde{\mathcal{S}}} \in \mathbb{R}^{|\widetilde{\mathcal{S}}| \times d}$ and $\mathbf{E}_{\widetilde{\mathcal{R}}} \in \mathbb{R}^{|\widetilde{\mathcal{R}}| \times d}$ represent the embeddings of the \srcandrec codes, respectively.

After obtaining the enhanced user \srcandrec preference embeddings, $\mathbf{e}_u^{s}$ and $\mathbf{e}_u^{r}$, through propagation over the graphs $\mathcal{G}_s$ and $\mathcal{G}_r$, respectively, 
we align the two embeddings to better capture user-level similarity, which facilitates more effective information transfer during message passing.
Moreover, the aligned embeddings are also utilized in subsequent downstream tasks. Specifically, we adopt contrastive learning and compute the following InfoNCE loss:
\begin{equation}
\label{eq:user_align}
\begin{aligned}
    \mathcal{L}_{\mathrm{U\text{-}CL}}=
    - & \left[ \mathrm{log}\frac{\mathrm{exp}(\mathrm{sim}(\mathbf{e}_u^s,\mathbf{e}_u^r)/\tau_2)}{\sum_{u^{-} \in \mathcal{U}_{\mathrm{neg}}}\mathrm{exp}(\mathrm{sim}(\mathbf{e}_u^s,\mathbf{e}_{u^{-}}^r)/\tau_2)} \right.\\
    & \left. +~~\mathrm{log}\frac{\mathrm{exp}(\mathrm{sim}(\mathbf{e}_u^s,\mathbf{e}_u^r)/\tau_2)}{\sum_{u^{-} \in \mathcal{U}_\mathrm{neg}}\mathrm{exp}(\mathrm{sim}(\mathbf{e}_{u^{-}}^s,\mathbf{e}_u^r)/\tau_2)} \right],
\end{aligned}
\end{equation}
where $\tau_2$ is a learnable temperature coefficient and $\mathcal{U}_{\mathrm{neg}}$ is the set of in-batch negative users.
After message passing, we retrieve the embeddings of the user's \srcandrec code sequences, $\tilde{s}$ and $\tilde{r}$, by performing a lookup on the embedding tables $\mathbf{E}_{\widetilde{\mathcal{S}}}$ and $\mathbf{E}_{\widetilde{\mathcal{R}}}$, as learned from Eq.~\eqref{eq:user_code} in \S~\ref{sec:user_quant}, as follows:
\begin{equation}
\label{eq:code_emb}
\begin{aligned}
    \mathbf{E}_{\tilde{s}}=[\mathbf{e}_{s_1},\mathbf{e}_{s_2},\ldots,\mathbf{e}_{s_L}]^{\mathsf{T}} \in \mathbb{R}^{L \times d}, ~~
    \mathbf{E}_{\tilde{r}}=[\mathbf{e}_{r_1},\mathbf{e}_{r_2},\ldots,\mathbf{e}_{r_L}]^{\mathsf{T}} \in \mathbb{R}^{L \times d}.
\end{aligned}
\end{equation}
The aligned user embeddings $\mathbf{e}_u^s$ and $\mathbf{e}_u^r$, as well as the \srcandrec code sequence embeddings 
$\mathbf{E}_{\tilde{s}}$ and $\mathbf{E}_{\tilde{r}}$, 
are subsequently utilized in downstream modeling.

\subsubsection{History Modeling}
\label{sec:history_model}
We first obtain the embeddings of the user's \srcandrec histories via the lookup operation. Specifically, the embedding of the recommendation history is obtained by concatenating the embeddings of the constituent items:
\begin{equation*}
\mathbf{E}_{r}=[\mathbf{e}_{i_1},\mathbf{e}_{i_2},\ldots,\mathbf{e}_{i_{N_r}}]^{\mathsf{T}} \in \mathbb{R}^{N_r \times d}.
\end{equation*}
For the search history, the embedding of each record is computed by summing the embedding of the query and the mean-pooled embedding of its associated clicked items. The overall embedding of the search history is formulated as:
\begin{small}
\begin{equation*}
\mathbf{E}_{s}=[\mathbf{e}_{q_1} + \mathrm{M}(\mathcal{I}_{q_1}),\mathbf{e}_{q_2} + \mathrm{M}(\mathcal{I}_{q_2}),\ldots,\mathbf{e}_{q_{N_s}} + \mathrm{M}(\mathcal{I}_{q_{N_s}})]^{\mathsf{T}} \in \mathbb{R}^{N_s \times d},
\end{equation*}
\end{small}
where $\mathrm{M}(\mathcal{I}_{q_k}) = \mathrm{MEAN}(\mathbf{e}_{i_1},\mathbf{e}_{i_2},\ldots,\mathbf{e}_{i_{N_{q_k}}})$ denotes the mean of the embeddings of items clicked in response to query $q_k$.

To capture the sequential dependencies within user behavior sequences, we introduce position embeddings $\mathbf{P}_s \in \mathbb{R}^{N_s \times d}$ and $\mathbf{P}_r \in \mathbb{R}^{N_r \times d}$ for the \srcandrec histories, respectively.
The final representations of the \srcandrec histories are computed as follows: 
\begin{small}
\begin{equation*}
    \widehat{\mathbf{E}}_s = \mathbf{E}_{s} + \mathbf{P}_s, \quad
    \widehat{\mathbf{E}}_r = \mathbf{E}_{r} + \mathbf{P}_r.
\end{equation*}
\end{small}

To further model the contextual representations of user \srcandrec histories, we encode them separately using two Transformer~\cite{vaswani2017attention} encoders, each consisting of a Multi-Head Self-Attention (MSA) layer followed by a Feed-Forward Network (FFN). 
The historical embeddings serve as the query, key, and value in the MSA mechanism.
The encoding process is formulated as:
\begin{small}
\begin{equation*}
\mathbf{H}_s = \mathrm{FFN}_s(\mathrm{MSA}_s(\widehat{\mathbf{E}}_s,\widehat{\mathbf{E}}_s,\widehat{\mathbf{E}}_s)),~~
\mathbf{H}_r = \mathrm{FFN}_r(\mathrm{MSA}_r(\widehat{\mathbf{E}}_r,\widehat{\mathbf{E}}_r,\widehat{\mathbf{E}}_r)),
\end{equation*}
\end{small}
where $\mathbf{H}_s \in \mathbb{R}^{N_s \times d}$ and $\mathbf{H}_r \in \mathbb{R}^{N_r \times d}$ denote the contextualized embeddings of the \srcandrec histories, respectively.

$\mathbf{H}_s$ and $\mathbf{H}_r$ encode the user's interests reflected in their \srcandrec histories, respectively. In contrast, the code embeddings 
$\mathbf{E}_{\tilde{s}}$ and $\mathbf{E}_{\tilde{r}}$
derived in Eq.~\eqref{eq:code_emb} capture user preferences enhanced by collaborative relationships among users. We fuse these two types of representations to obtain enriched representations of the user's \srcandrec histories.
Specifically, we employ Multi-Head Cross-Attention (MCA), where the history representations serve as queries, and the corresponding code embeddings act as keys and values. The fusion process is computed as follows:
\begin{equation}
\label{eq:code_his_fusion}
\begin{aligned}
\mathbf{F}_s &= \mathrm{MSA}_s(\mathbf{H}_s,\mathbf{H}_s,\mathbf{H}_s),~~\mathbf{W}_s=\mathrm{FFN}_s(\mathrm{MCA}_s(\mathbf{F}_s, \mathbf{E}_{\tilde{s}}, \mathbf{E}_{\tilde{s}})),\\
\mathbf{F}_r &= \mathrm{MSA}_r(\mathbf{H}_r,\mathbf{H}_r,\mathbf{H}_r),~~\mathbf{W}_r=\mathrm{FFN}_r(\mathrm{MCA}_r(\mathbf{F}_r, \mathbf{E}_{\tilde{r}}, \mathbf{E}_{\tilde{r}})),\\
\end{aligned}
\end{equation}
where $\mathbf{W}_s \in \mathbb{R}^{N_s \times d}$ and $\mathbf{W}_r \in \mathbb{R}^{N_r \times d}$ denote the final contextually enriched representations for the \srcandrec histories, respectively.

To enable more effective fusion, we first align the history embeddings with the code embeddings. Taking the search history as an example, we compute the mean of the search history embeddings and the search code sequence embeddings to obtain $\mathbf{h}_s = \mathrm{MEAN}(\mathbf{H}_s) \in \mathbb{R}^d$ and 
$\mathbf{e}_{\tilde{s}} = \mathrm{MEAN}(\mathbf{E}_{\tilde{s}}) \in \mathbb{R}^d$, 
respectively. 
Then we employ contrastive learning and computer the following loss:
\begin{equation}
\label{eq:code_his_align_src}
\begin{aligned}
\mathcal{L}_{\mathrm{S\text{-}CL}}=
- & \left[ \mathrm{log}\frac{\mathrm{exp}(\mathrm{sim}(\mathbf{h}_s,\mathbf{e}_{\tilde{s}})/\tau_3)}{\sum_{\mathbf{e}_{\tilde{s}}^{-} \in \mathcal{E}_{\tilde{s}}^{\mathrm{neg}}}\mathrm{exp}(\mathrm{sim}(\mathbf{h}_s,\mathbf{e}_{\tilde{s}}^{-})/\tau_3)} \right.\\
& \left. +~~\mathrm{log}\frac{\mathrm{exp}(\mathrm{sim}(\mathbf{h}_s,\mathbf{e}_{\tilde{s}})/\tau_3)}{\sum_{\mathbf{h}_s^{-} \in \mathcal{H}_{s}^\mathrm{neg}}\mathrm{exp}(\mathrm{sim}(\mathbf{h}_s^{-},\mathbf{e}_{\tilde{s}})/\tau_3)} \right],
\end{aligned}
\end{equation}
where $\tau_3$ is a learnable temperature coefficient, 
$\mathcal{E}_{\tilde{s}}^{\mathrm{neg}}$ 
and $\mathcal{H}_{s}^\mathrm{neg}$ are in-batch negative samples.
Similarly, we can get the contrastive loss $\mathcal{L}_{\mathrm{R\text{-}CL}}$ for recommendation history and code sequence. The total contrastive loss is formulated as follows:
\begin{equation}
\label{eq:code_his_align}
    \mathcal{L}_{\mathrm{His\text{-}CL}}=\mathcal{L}_{\mathrm{S\text{-}CL}} + \mathcal{L}_{\mathrm{R\text{-}CL}}.
\end{equation}

After obtaining the representations of the user's \srcandrec histories, $\mathbf{W}_s$ and $\mathbf{W}_r$, we perform history pooling based on the similarity between each historical behavior and the next candidate item. Specifically, we apply a Self-Attention (SA) mechanism as follows:
\begin{equation}
\label{eq:history_pool}
    \mathbf{w}_s = \mathrm{SA}(\mathbf{e}_{i_{T+1}},\mathbf{W}_s,\mathbf{W}_s),\quad
    \mathbf{w}_r = \mathrm{SA}(\mathbf{e}_{i_{T+1}},\mathbf{W}_r,\mathbf{W}_r),
\end{equation}
where $\mathbf{w}_s, \mathbf{w}_r \in \mathbb{R}^d$ are the aggregated representations of the \srcandrec histories, respectively. Here, the embedding of the next candidate item $\mathbf{e}_{i_{T+1}}$ serves as the query, while the \srcandrec history representations act as the key and value for the attention computation.

\subsection{Model Prediction and Training}

\subsubsection{Prediction}
Finally, we concatenate the user’s \srcandrec representations, $\mathbf{e}_u^s$ and $\mathbf{e}_u^r$, obtained from Eq.~\eqref{eq:lightgcn_final}, along with the historical representations $\mathbf{w}_s$ and $\mathbf{w}_r$ derived from Eq.~\eqref{eq:history_pool}, and the embedding of the next candidate item $\mathbf{e}_{i_{T+1}}$. The concatenated vector is then fed into a multi-layer perceptron (MLP) to predict the user’s preference for the next item:
\begin{equation}
    \hat{y}_{u,i_{T+1}} = \mathrm{MLP}(\mathrm{CONCAT}(\mathbf{e}_u^s,\mathbf{e}_u^r,\mathbf{w}_s,\mathbf{w}_r,\mathbf{e}_{i_{T+1}})),
\end{equation}
where $\hat{y}_{u,i_{T+1}}$ is the predicted preference score. $\mathrm{CONCAT}(\cdot)$ denotes the concatenation operation.

\subsubsection{Training}
Following previous works~\cite{DIN,SESRec,UniSAR}, 
we adopt the binary cross-entropy loss to optimize our recommendation model: 
\begin{small}
\begin{equation}
\label{eq:click_loss}
    \mathcal{L}_{\mathrm{rec}} = -\frac{1}{|\mathcal{D}|} \sum_{(u,i_{T+1}) \in \mathcal{D}} y_{u,i_{T+1}}\mathrm{log}(\hat{y}_{u,i_{T+1}}) + (1-y_{u,i_{T+1}})\mathrm{log}(1-\hat{y}_{u,i_{T+1}}),  \\
\end{equation}
\end{small}
where $\mathcal{D}$ denotes the set of user-item interaction pairs used for training. Here, $y_{u,i_{T+1}} \in \{0,1\}$ is the ground-truth label indicating whether user $u$ interacts with item $i_{T+1}$.

Finally, the overall training loss of our model combines the recommendation loss defined in Eq.~\eqref{eq:click_loss} with two auxiliary contrastive losses introduced in Eq.~\eqref{eq:user_align} and Eq.~\eqref{eq:code_his_align}. Additionally, we incorporate an $L_2$ regularization term to prevent overfitting. The total loss is formulated as:
\begin{equation}
\label{eq:total_loss}
    \mathcal{L}_{\mathrm{Total}} = \mathcal{L}_{\mathrm{rec}} + \lambda_{\mathrm{U\text{-}CL}}\mathcal{L}_{\mathrm{U\text{-}CL}} + \lambda_{\mathrm{His\text{-}CL}}\mathcal{L}_{\mathrm{His\text{-}CL}} + \lambda_{\mathrm{Reg}}||\Theta||^2,
\end{equation}
where $\lambda_{\mathrm{U\text{-}CL}}$, $\lambda_{\mathrm{His\text{-}CL}}$, and $\lambda_{\mathrm{Reg}}$ are hyper-parameters that control the contributions of the user-level alignment loss, history-level alignment loss, and regularization term, respectively. Here, $||\Theta||^2$ represents the $L_2$ norm of the model parameters $\Theta$, which helps to regularize the model and improve its generalization ability.

\subsection{Discussion}

\paratitle{Computational Efficiency and Complexity.}
Our method leverages LLM inference solely for summarizing user preferences, which can be performed offline. As a result, the online serving phase only requires the recommendation model, ensuring high efficiency.
Regarding the user-code graph, take $\mathcal{G}_s$ (described in \S~\ref{sec:graph_construct}) as an example. It consists of $|\mathcal{U}| + |\widetilde{\mathcal{S}}|$ nodes, where $|\mathcal{U}|$ denotes the number of users and $|\widetilde{\mathcal{S}}|$ the number of search codes. The number of codes $|\widetilde{\mathcal{S}}|$ is at most $L \times N_c$, where $L$ is the number of codebooks used in RQ-VAE (\S~\ref{sec:user_quant}) and $N_c$ is the size of each codebook. In practice, both $L \ll |\mathcal{U}|$ and $N_c \ll |\mathcal{U}|$, so the resulting graph remains lightweight.
This structure is significantly more efficient than prior graph-based recommendation models~\cite{he2020lightgcn,wu2021self,yu2022graph}, where the node count is $|\mathcal{U}| + |\mathcal{I}|$, with $|\mathcal{I}|$ (the number of items) typically much larger than both $L$ and $N_c$ ($|\mathcal{I}| \gg L$ and $|\mathcal{I}| \gg N_c$).

\paratitle{Comparison with Existing Methods.}
In contrast to existing search-enhanced recommendation models, our method explicitly targets the challenge of sparse search interactions by constructing the user-code graphs. Through message passing on the graphs, user embeddings with rich search interactions are leveraged to enhance those of users with sparse behaviors. This design leads to more substantial performance gains, especially for users with limited search histories.

Compared to GNN-based graph recommendation models~\cite{he2020lightgcn,wu2021self,yu2022graph}, which commonly construct user-item graphs to capture collaborative filtering signals, our approach instead builds the user-code graphs aimed at enhancing user representations. This graph structure facilitates more effective user-user information sharing via shared discrete codes, offering a novel perspective on graph-based representation learning.

\begin{table}[t]
    \caption{
    Dataset statistics used in this paper. ``S'' and ``R'' represent search and recommendation, respectively.}
    \vspace{-8px}
    \center
     \resizebox{.98\columnwidth}{!}{
        \begin{tabular}{lccccc}
        \toprule
        Dataset & \#Users & \#Items & \#Queries & \#Interaction-S &\#Interaction-R  \\
        \midrule
        CDs &75,258 &64,443 &671  &852,889 &1,097,592 \\
        Electronics &192,403 &63,001 &982 &1,280,465 &1,689,188 \\
        Qilin &15,482 &1,983,938 &44,820 &969,866 &1,438,435 \\
        \bottomrule
        \end{tabular}}
    \label{tab:dataStatistics}   
   \vspace{-0.5cm}
\end{table}

\begin{table*}[t]
\centering
\caption{
Overall recommendation performance comparison of different methods on all datasets.
H@$k$ and N@$k$ denote HR@$k$ and NDCG@$k$, respectively.
The best and second-best results are highlighted in bold and underlined fonts, respectively.
``*'' denotes that the improvement over the second-best method is statistically significant ($t$-test, $p$-value $< 0.05$).
}
\vspace{-8px}
\label{tab:rec_result}
\resizebox{0.95\linewidth}{!}{
\begin{tabular}{
lc
ccccc
ccccc
ccccc
}
\toprule
\multicolumn{1}{c}{\multirow{2}{*}{Category}} & 
\multicolumn{1}{c}{\multirow{2}{*}{Methods}} & 
\multicolumn{5}{c}{CDs} & 
\multicolumn{5}{c}{Electronics} & 
\multicolumn{5}{c}{Qilin} \\
\cmidrule(l){3-7} \cmidrule(l){8-12} \cmidrule(l){13-17} 
\multicolumn{1}{c}{}  & \multicolumn{1}{c}{}  
&H@1 &H@5 &H@10 &N@5 &N@10
&H@1 &H@5 &H@10 &N@5 &N@10
&H@1 &H@5 &H@10 &N@5 &N@10 \\
\midrule
\multirow{9}*{Recommendation} 
&LightGCN &0.0911 &0.3285 &0.4963 &0.2103 &0.2645 &0.0277 &0.1025 &0.1707 &0.0648 &0.0867 &0.0624 &0.2436 &0.3820 &0.1530 &0.1976\\
&SGL &0.1431 &0.3660 &0.4929 &0.2579 &0.2988 &0.0304 &0.1081 &0.1816 &0.0690 &0.0926 &0.0780 &0.2566 &0.3854 &0.1684 &0.2099\\
&SimGCL &0.1651 &0.3874 &0.5141 &0.2799 &0.3207 &0.0395 &0.1261 &0.2061 &0.0827 &0.1084 &0.0751 &0.2607 &0.3889 &0.1684 &0.2097\\
&GRU4Rec &0.1360 &0.4191 &0.5854 &0.2806 &0.3344 &0.0579 &0.2005 &0.3144 &0.1295 &0.1661 &0.1295 &0.3443 &0.4855 &0.2395 &0.2850 \\
&SASRec &0.1621 &0.4134 &0.5638 &0.2907 &0.3393 &0.1019 &0.2188 &0.3121 &0.1608 &0.1907 &0.1334 &0.3449 &0.4768 &0.2411 &0.2836 \\
&BERT4Rec &0.1692 &0.4226 &0.5750 &0.2993 &0.3485 &0.1031 &0.2242 &0.3224 &0.1642 &0.1957 &0.1319 &0.3472 &0.4832 &0.2426 &0.2864 \\
&CL4SRec &0.1834 &0.4570 &0.6084 &0.3240 &0.3730 &0.1052 &0.2308 &0.3299 &0.1684 &0.2003 &0.1363 &0.3489 &0.4848 &0.2455 &0.2894 \\
&KAR &0.1922 &0.4937 &0.6394 &0.3483 &0.3955 &0.0958 &0.2513 &0.3629 &0.1750 &0.2108 &0.1140 &0.3200 &0.4551 &0.2188 &0.2625 \\
&LLM-ESR &0.2079 &0.5104 &0.6610 &0.3648 &0.4136 &0.1055 &0.2560 &0.3672 &0.1817 &0.2175 &0.1422 &0.3599 &0.4932 &0.2532 &0.2963 \\
\hline
\multirow{8}*{\makecell[l]{Search Enhanced \\ Recommendation}}
&NRHUB &0.1454 &0.4243 &0.5825 &0.2885 &0.3397 &0.0533 &0.1820 &0.2889 &0.1179 &0.1522 &0.1389 &0.3543 &0.4829 &0.2499 &0.2913 \\
&Query-SeqRec &0.1832 &0.4537 &0.6066 &0.3219 &0.3713 &0.1009 &0.2219 &0.3205 &0.1619 &0.1935 &0.1299 &0.3473 &0.4824 &0.2412 &0.2847 \\
&JSR &0.1808 &0.4346 &0.5807 &0.3113 &0.3586 &\underline{0.1090} &0.2289 &0.3246 &0.1694 &0.2001 &0.1445 &0.3711 &0.5077 &0.2608 &0.3048 \\
&USER &0.1904 &0.4929 &0.6465 &0.3465 &0.3963 &0.0672 &0.2146 &0.3270 &0.1415 &0.1776 &0.1549 &0.3820 &\underline{0.5199} &0.2715 &0.3161 \\
&SESRec &0.2019 &0.5059 &0.6494 &0.3595 &0.4060 &0.0790 &0.2403 &0.3572 &0.1606 &0.1982 &0.1535 &0.3694 &0.4904 &0.2647 &0.3037 \\
&UnifiedSSR &0.2079 &0.4928 &0.6359 &0.3549 &0.4012 &0.1066 &0.2343 &0.3320 &0.1711 &0.2025 &0.1412 &0.3595 &0.4957 &0.2532 &0.2972 \\
&UniSAR &\underline{0.2219} &\underline{0.5249} &\underline{0.6712} &\underline{0.3797} &\underline{0.4271} &{0.0996} &\underline{0.2633} &\underline{0.3757} &\underline{0.1829} &\underline{0.2191} &\underline{0.1616} &\underline{0.3835} &{0.5099} &\underline{0.2767} &\underline{0.3174} \\
&\textbf{\ourname} &\textbf{0.2505}* &\textbf{0.5459}* &\textbf{0.6825}* &\textbf{0.4045}* &\textbf{0.4487}* &\textbf{0.1205}* &\textbf{0.2739}* &\textbf{0.3788}* &\textbf{0.1989}* &\textbf{0.2327}* &\textbf{0.1812}* &\textbf{0.4062}* &\textbf{0.5335}* &\textbf{0.2988}* &\textbf{0.3400}* \\
\bottomrule
\end{tabular} 
}
\vspace{-0.3cm}
\end{table*}

\section{Experiments}
We conducted extensive experiments to evaluate the performance of \ourname.

\subsection{Experimental Setup}

\subsubsection{Dataset}
Since \ourname relies on both users' \srcandrec interaction data, as well as the textual information of items, we conduct experiments on the following publicly available datasets.
The statistics of these datasets are summarized in Table~\ref{tab:dataStatistics}.

\textbf{Amazon}\footnote{\url{https://cseweb.ucsd.edu/~jmcauley/datasets/amazon/links.html},~\url{https://github.com/QingyaoAi/Amazon-Product-Search-Datasets}.}~\cite{amazon_dataset,amazon_dataset2}:
We adopted a widely used semi-synthetic dataset. Following previous studies~\cite{ai2017learning, ai2019zero, SESRec, UniSAR}, we generated synthetic search behaviors based on an existing recommendation dataset.
We used the ``CDs and Vinyl'' and ``Electronics'' subsets and selected their five-core versions, ensuring that each user and item has at least five interactions.
\footnote{
Due to the lack of textual information for approximately 70\% of items in the ``Kindle Store'' subset used in previous works~\cite{SESRec, UniSAR}, we instead used the ``CDs and Vinyl'' and ``Electronics'' subsets, where fewer than 1\% of items lack text.
}

\textbf{Qilin}~\cite{chen2025qilin}:
The dataset is collected from Xiaohongshu\footnote{\url{https://www.xiaohongshu.com}.}, a well-known lifestyle search engine in China with over 300 million monthly active users.
It contains user behavior data from both \srcandrec scenarios, as well as multimodal information for all items. In this work, we utilize only the textual information.

Following previous works~\cite{BERT4REC,SESRec,UniSAR}, we applied the leave-one-out strategy to split all the dataset into training, validation, and test~sets.

\subsubsection{Baselines}
We compare \ourname with two categories of baselines to comprehensively evaluate its effectiveness:

(1)~\emph{Recommendation}:
\textbf{LightGCN}~\cite{he2020lightgcn} is a simple graph-based model that captures user-item interactions via neighborhood aggregation;
\textbf{SGL}~\cite{wu2021self} is a graph-based recommendation model that enhances representations through self-supervised contrastive learning;
\textbf{SimGCL}~\cite{yu2022graph} improves recommendation by injecting noise into embeddings for contrastive learning;
\textbf{GRU4Rec}~\cite{GRU4REC} models user interaction histories using gated recurrent units (GRUs);
\textbf{SASRec}~\cite{SASREC} is a sequential recommendation model based on a unidirectional Transformer architecture;
\textbf{BERT4Rec}~\cite{BERT4REC} employs a cloze-style objective with a bidirectional Transformer for sequential recommendation;
\textbf{CL4SRec}~\cite{xie2022contrastive} designs a contrastive learning objective to alleviate data sparsity.
\textbf{KAR}~\cite{xi2024towards} enhances recommendation by injecting LLM-generated user reasoning and item knowledge as precomputed vectors;
\textbf{LLM-ESR}~\cite{liu2024llm} enhances sequential recommenders by combining LLM semantics and collaborative signals to tackle long-tail user and item challenges.

(2)~\emph{Search enhanced recommendation}:
\textbf{NRHUB}~\cite{NRHUB} is a news recommendation model that leverages heterogeneous user behaviors;
\textbf{Query-SeqRec}~\cite{Query_SeqRec} is a query-aware sequential model that incorporates user queries into behavioral sequences using Transformers;
\textbf{JSR}~\cite{JSR} is a general framework that optimizes a joint loss over multiple tasks;
\textbf{USER}~\cite{USER} integrates user behaviors from both \srcandrec into a unified heterogeneous behavior sequence;
\textbf{SESRec}~\cite{SESRec} employs contrastive learning to learn disentangled search representations for recommendation;
\textbf{UnifiedSSR}~\cite{xie2024unifiedssr} jointly models user behavior histories across both \srcandrec scenarios;
\textbf{UniSAR}~\cite{UniSAR} models user transition behaviors between \srcandrec by employing transformers with distinct masking strategies and contrastive learning objectives.

\subsubsection{Evaluation}
Following previous studies~\cite{BERT4REC, SESRec,UniSAR}, we adopt \textit{Hit Ratio}~(HR) and \textit{Normalized Discounted Cumulative Gain}~(NDCG) as our evaluation metrics.
We report HR at top $\{1, 5, 10\}$ ranks and NDCG at top $\{5, 10\}$ ranks.
Following the standard evaluation protocol~\cite{SASREC, SESRec,UniSAR}, each ground-truth item is paired with 99 randomly sampled negative items with which the user has no prior interactions to form the candidate list.

\subsubsection{Implementation Details}
User-Code Graph Construction (\S~\ref{sec:user_code_graph}):
We use the LLM DeepSeek-R1-Distill-Qwen-7B\footnote{\url{https://huggingface.co/deepseek-ai/DeepSeek-R1-Distill-Qwen-7B}.}~\cite{deepseek_r1} to summarize user \srcandrec preferences, which are then embedded via BGE-M3\footnote{\url{https://huggingface.co/BAAI/bge-m3}.}~\cite{bge-m3}. The RQ-VAE (\S~\ref{sec:user_quant}) uses $L=4$ codebooks with $N_c=256$ codes each and code dimension $d_l=32$. We set $\lambda_{\mathrm{RQ}}=1.0$ (Eq,~\eqref{eq:rq_total_loss}) and train the quantization model for 500 epochs using Adam~\cite{kingma2014adam} with a batch size of 1024 and learning rate 1e-3. The temperature $\tau_1$ (Eq.~\eqref{eq:rq_user_align}) is 0.1, and the contrastive loss weight $\lambda_{\mathrm{RQ\text{-}CL}}$ (Eq.~\eqref{eq:rq_total_loss}) is tuned from \{1e-4, 1e-3, 1e-2, 1e-1, 1\}.

Search Enhanced Recommendation Modeling (\S~\ref{sec:sequential_model}):
Each baseline was tuned per dataset based on the original paper settings.
For our model, embedding dimension $d$ is 64 for CDs/Electronics and 32 for Qilin. The max history length is 20 for CDs/Qilin and 10 for Electronics. We use 2 LightGCN layers ($K=2$). Temperatures $\tau_2$ (Eq.~\eqref{eq:user_align}) and $\tau_3$ (Eq.~\eqref{eq:code_his_align_src}) are set to 0.1. Loss weights $\lambda_{\mathrm{U\text{-}CL}}$ and $\lambda_{\mathrm{His\text{-}CL}}$ (Eq.~\eqref{eq:total_loss}) are tuned over \{1e-4, 1e-3, 1e-2, 1e-1, 1\}. All models are trained for up to 100 epochs with Adam~\cite{kingma2014adam}, using a batch size of 1024 and early stopping. The learning rate is searched from \{1e-3, 1e-4, 1e-5\}, and $\lambda_{\mathrm{Reg}}$ (Eq.~\eqref{eq:total_loss}) is tuned over \{1e-5, 1e-6, 1e-7\}.

\begin{table}[t!]
\small
\caption{
Ablation study conducted on the Qilin dataset, where ``w/o'' indicates that the corresponding module has been removed from \ourname. ``U-C Graph'' denotes the user-code~graph.  
}
\vspace{-8px}
\label{tab:ablation_result}
\renewcommand{\arraystretch}{1.2}
\resizebox{0.95\columnwidth}{!}{
\begin{tabular}
{l
ccccc}
\toprule
Model 
&H@1 &H@5 &H@10 
&N@5 &N@10  \\ 
\midrule
\textbf{\ourname} &\textbf{0.1812} &\textbf{0.4062} &\textbf{0.5335} &\textbf{0.2988} &\textbf{0.3400} \\
\hdashline
w/o $\mathcal{L}_{\mathrm{RQ\text{-}CL}}$ (Eq.~\eqref{eq:rq_user_align}) &0.1665 &0.3975 &0.5272 &0.2862 &0.3281 \\
\hdashline
w/o U-C Graph &0.1428 &0.3647 &0.4858 &0.2576 &0.2967 \\
w/o $\mathcal{L}_{\mathrm{U\text{-}CL}}$ (Eq.~\eqref{eq:user_align}) &0.1632 &0.3922 &0.5230 &0.2819 &0.3242 \\
\hdashline
w/o $\mathcal{L}_{\mathrm{His\text{-}CL}}$ (Eq.~\eqref{eq:code_his_align}) &0.1654 &0.3935 &0.5215 &0.2835 &0.3248 \\
w/o MCA &0.1763 &0.4032 &0.5295 &0.2933 &0.3358 \\
\bottomrule
\end{tabular}
} 
\vspace{-0.5cm}
\end{table}

\subsection{Overall Performance}
Table~\ref{tab:rec_result} presents the recommendation results on three datasets. From the results, we can observe the following:

\noindent\textbf{$\bullet $}~Firstly, it can be observed that compared to existing recommendation or search-enhanced recommendation models, \ourname achieves state-of-the-art results. 
This validates the effectiveness of \ourname in alleviating data sparsity by constructing the user-code graphs and performing message passing, thereby enhancing the representations of users with sparse search interactions using information from users with richer search behaviors.

\noindent\textbf{$\bullet $}~Secondly, we observe that search-enhanced recommendation models, such as \ourname and UniSAR, generally outperform traditional recommendation methods. However, models like NRHUB perform worse than traditional baselines in some cases, indicating that simply incorporating search features does not necessarily lead to improved performance. This highlights the need for dedicated designs to effectively learn representations of search~features.

\noindent\textbf{$\bullet $}~Thirdly, we also observe that graph-based models such as LightGCN underperform compared to sequential recommendation models like SASRec, highlighting the importance of leveraging users’ historical behaviors. 
Meanwhile, models leveraging LLMs, including \ourname, KAR, and LLM-ESR, achieve significant improvements over traditional recommendation methods, highlighting the effectiveness of incorporating LLMs into recommendation tasks.

\noindent\textbf{$\bullet $}~Finally, we compare the performance of the baselines and our model across user groups with varying numbers of search interactions, as shown in Figure~\ref{fig:intro_improve}. Due to space limitations, we report results on the Qilin dataset, comparing the traditional recommendation model SASRec, the search-enhanced model UniSAR, and our proposed model \ourname. As observed, UniSAR achieves larger improvements for users with rich search interactions, while \ourname further outperforms it for users with sparse search interactions. This further confirms the effectiveness of \ourname in alleviating the sparsity of search interactions.

\begin{figure}[t]
    \centering
    \includegraphics[width=0.98\columnwidth]{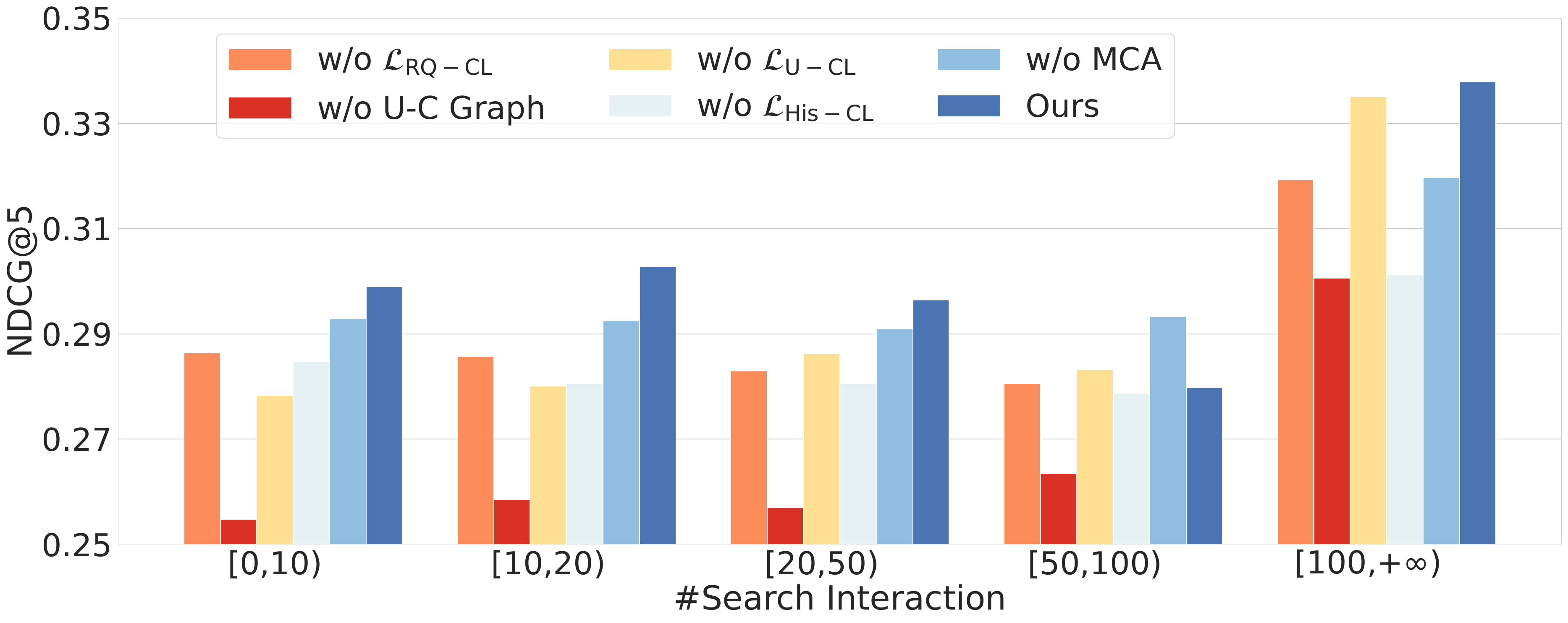}
   \vspace{-5px}
    \caption{
    Ablation study across user groups with varying numbers of search interactions
    }
    \label{fig:ablation_group}
    \vspace{-0.3cm}
\end{figure}

\subsection{Ablation Study}
Due to space limitations, we conduct ablation studies on Qilin, the dataset containing real user \srcandrec interactions, to evaluate the effectiveness of each module in \ourname.
The results are shown in Table~\ref{tab:ablation_result}.

\subsubsection{Effectiveness of User Alignment in Preference Quantization}
In \S~\ref{sec:user_quant}, we leverage contrastive learning to align the latent embeddings of users’ \srcandrec preferences, promoting the capture of user-level similarity. This alignment is enforced via the loss term $\mathcal{L}_{\mathrm{RQ\text{-}CL}}$ defined in Eq.~\eqref{eq:rq_user_align}. As demonstrated by the ``w/o $\mathcal{L}_{\mathrm{RQ\text{-}CL}}$'' setting in Table~\ref{tab:ablation_result}, removing this loss leads to a significant performance drop, underscoring its critical role. By ensuring better alignment before quantization, the generated codes more effectively reflect inter-user similarities, thereby improving downstream tasks.

\begin{figure*}[t]
     \centering
     \subfigure[$\mathbf{E}_{\mathcal{U}}^s$ and $\mathbf{E}_{\mathcal{U}}^r$ w/o $\mathcal{L}_{\mathrm{U\text{-}CL}}$.]{
        \label{fig:user_wo_ucl}
        \includegraphics[width=0.23\linewidth]{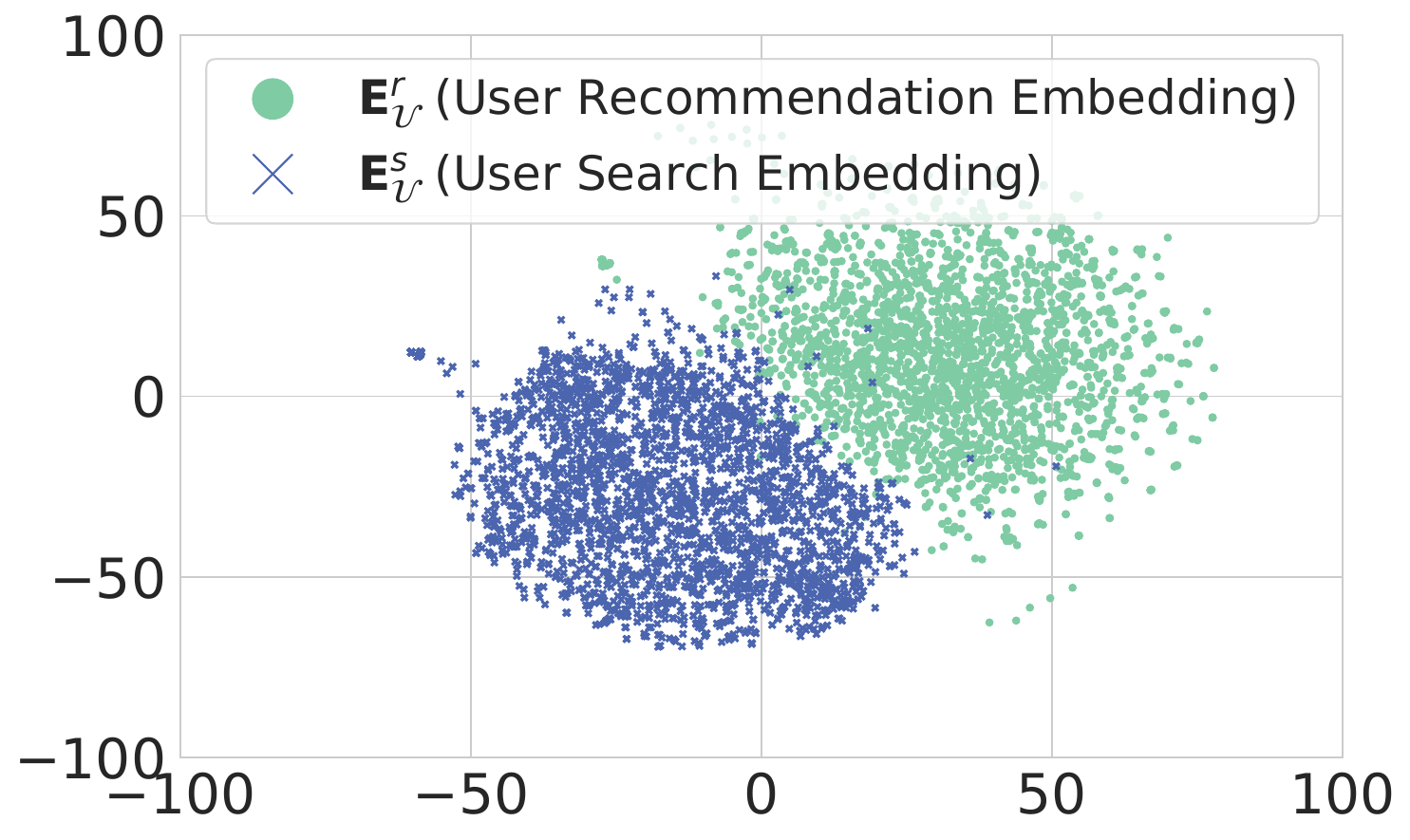}
     }
     \subfigure[$\mathbf{E}_{\mathcal{U}}^s$ and $\mathbf{E}_{\mathcal{U}}^r$ w/ $\mathcal{L}_{\mathrm{U\text{-}CL}}$.]{
        \label{fig:user_our}
        \includegraphics[width=0.23\linewidth]{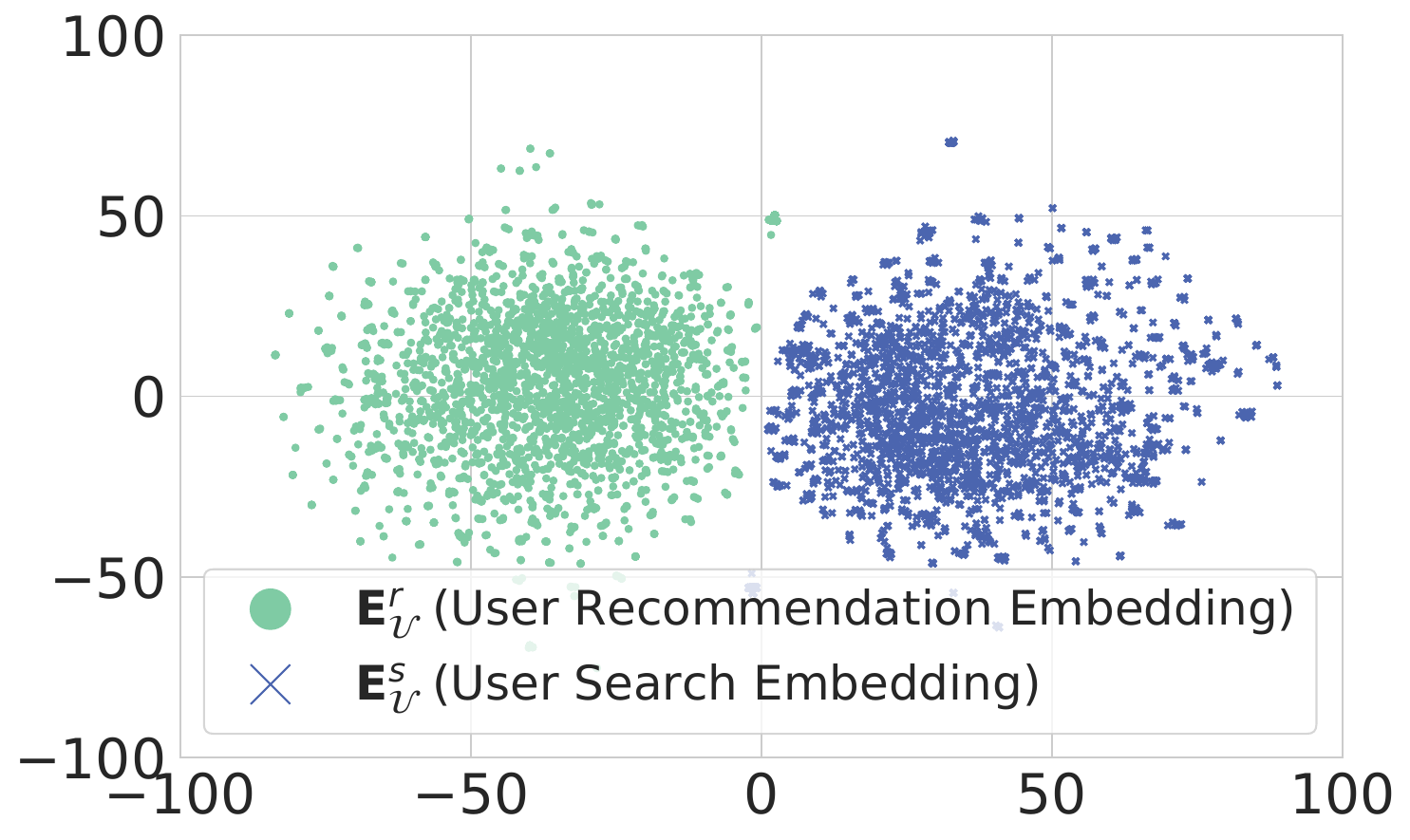}
     }
     \subfigure[$\mathbf{E}_{\widetilde{\mathcal{S}}}$ and $\mathbf{E}_{\widetilde{\mathcal{R}}}$ w/o $\mathcal{L}_{\mathrm{U\text{-}CL}}$.]{
        \label{fig:code_wo_ucl}
        \includegraphics[width=0.23\linewidth]{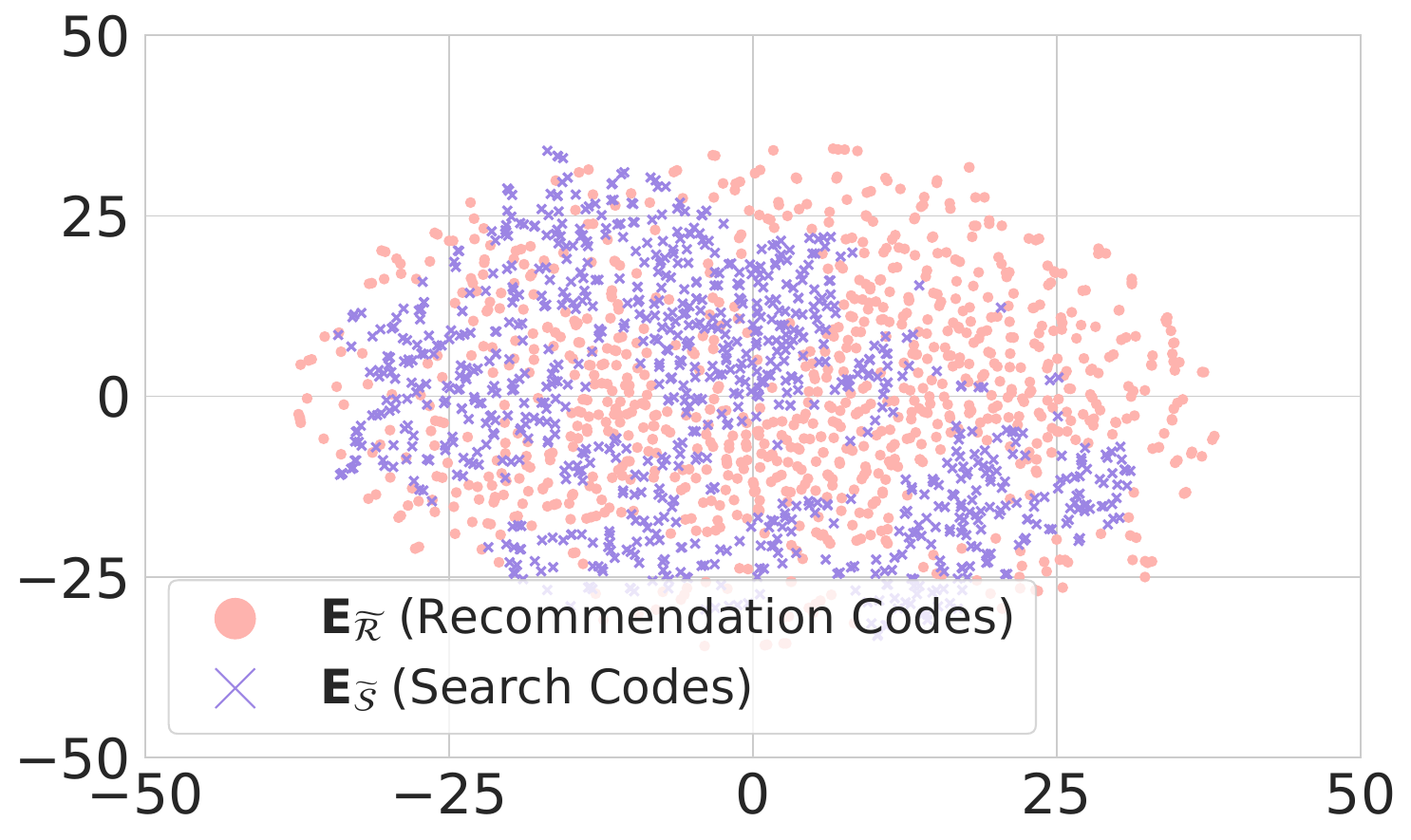}
     }
    \subfigure[$\mathbf{E}_{\widetilde{\mathcal{S}}}$ and $\mathbf{E}_{\widetilde{\mathcal{R}}}$ w/ $\mathcal{L}_{\mathrm{U\text{-}CL}}$.]{
        \label{fig:code_our}
        \includegraphics[width=0.23\linewidth]{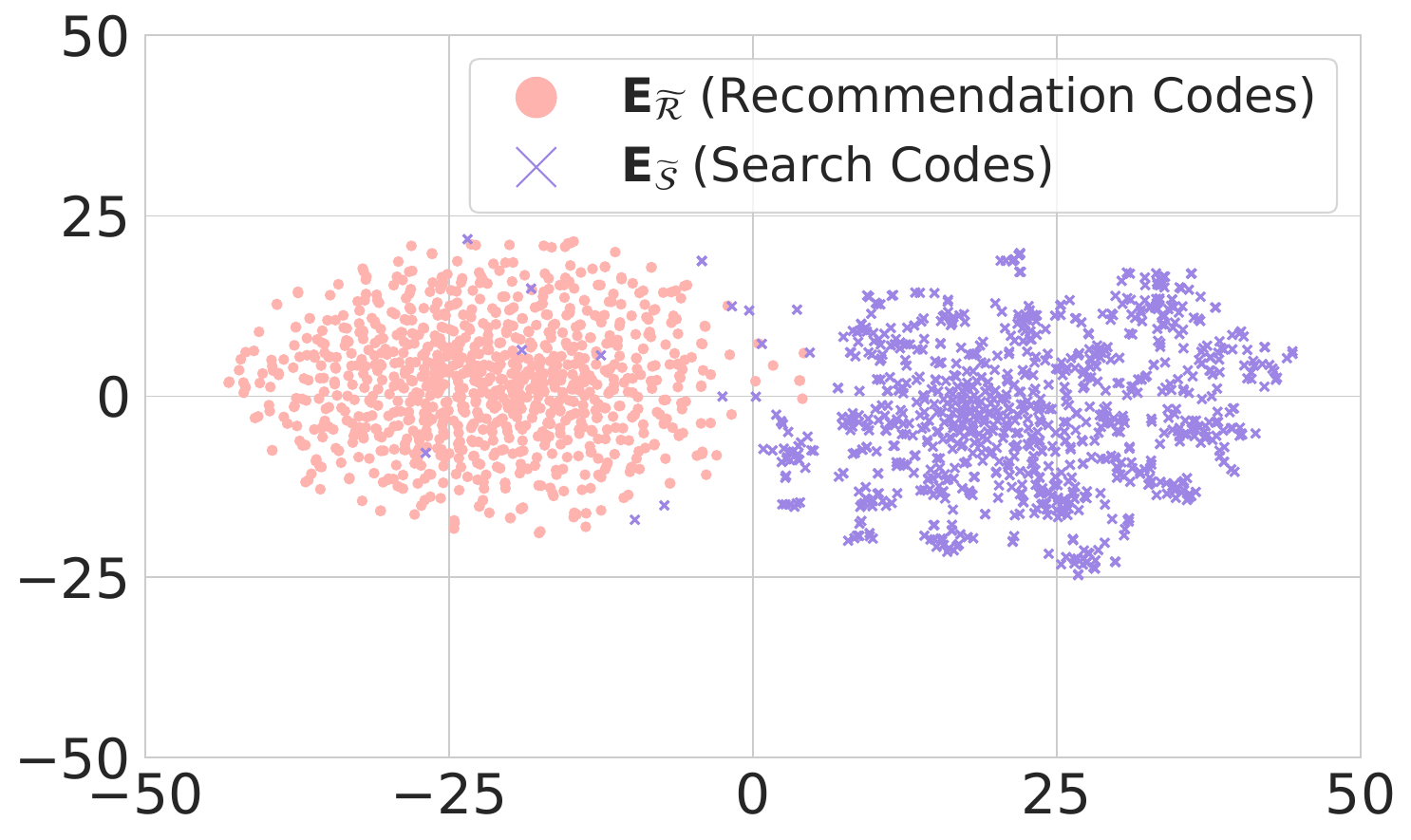}
     }
     \vspace{-8px}
     \caption{
     The t-SNE visualization of user \srcandrec embeddings $\mathbf{E}_{\mathcal{U}}^s$ and $\mathbf{E}_{\mathcal{U}}^r$, as well as code embeddings $\mathbf{E}_{\widetilde{\mathcal{S}}}$ and $\mathbf{E}_{\widetilde{\mathcal{R}}}$ (\S~\ref{sec:message_passing}), with and without the user alignment loss $\mathcal{L}_{\mathrm{U\text{-}CL}}$ (Eq.~\eqref{eq:user_align}).
     ``w/o'' and ``w/'' denote results without and with the alignment loss, respectively.
     }
     \label{fig:code_emb}
     \vspace{-0.5cm}
\end{figure*}

\begin{figure*}[t]
     \centering
     \subfigure[Performance of different $\lambda_{\mathrm{RQ\text{-}CL}}$ (Eq.~\eqref{eq:rq_total_loss})]{
        \label{fig:rq_cl_w}
        \includegraphics[width=0.31\linewidth]{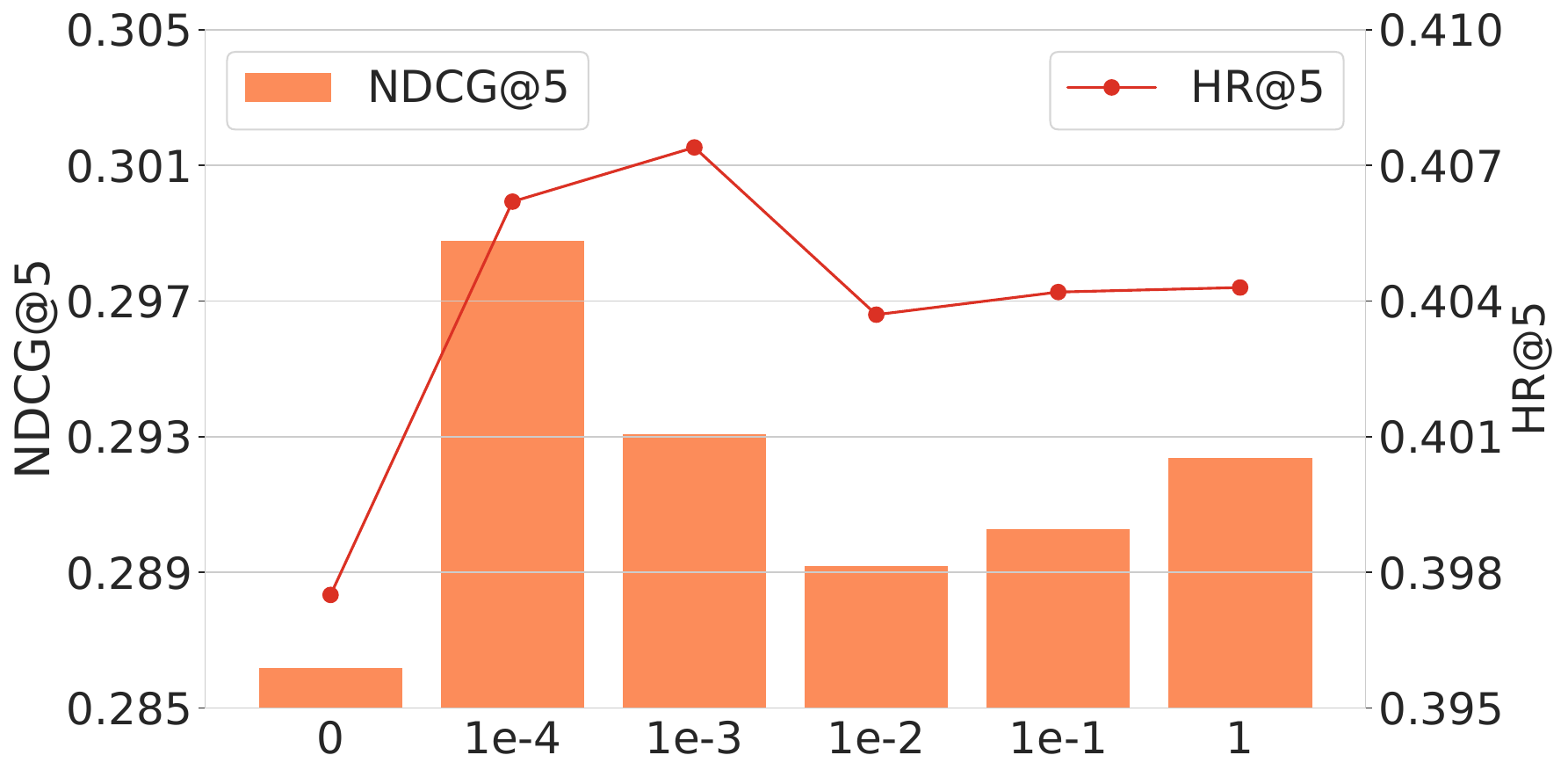}
     }
    \subfigure[Performance of different $\lambda_{\mathrm{U\text{-}CL}}$ (Eq.~\eqref{eq:total_loss})]{
        \label{fig:user_cl_w}
        \includegraphics[width=0.31\linewidth]{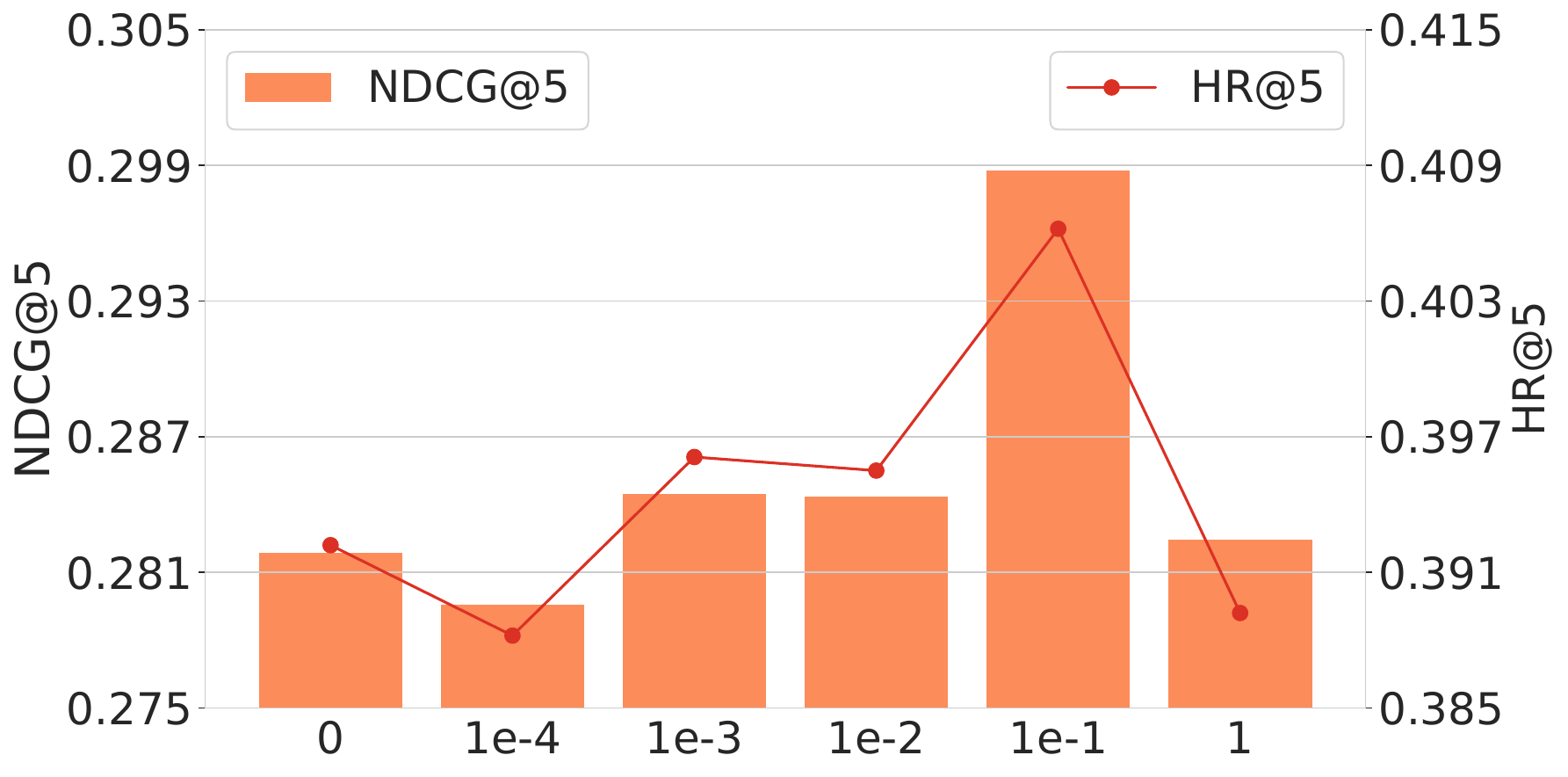}
     }
     \subfigure[Performance of different $\lambda_{\mathrm{His\text{-}CL}}$ (Eq.~\eqref{eq:total_loss})]{
        \label{fig:his_cl_w}
        \includegraphics[width=0.31\linewidth]{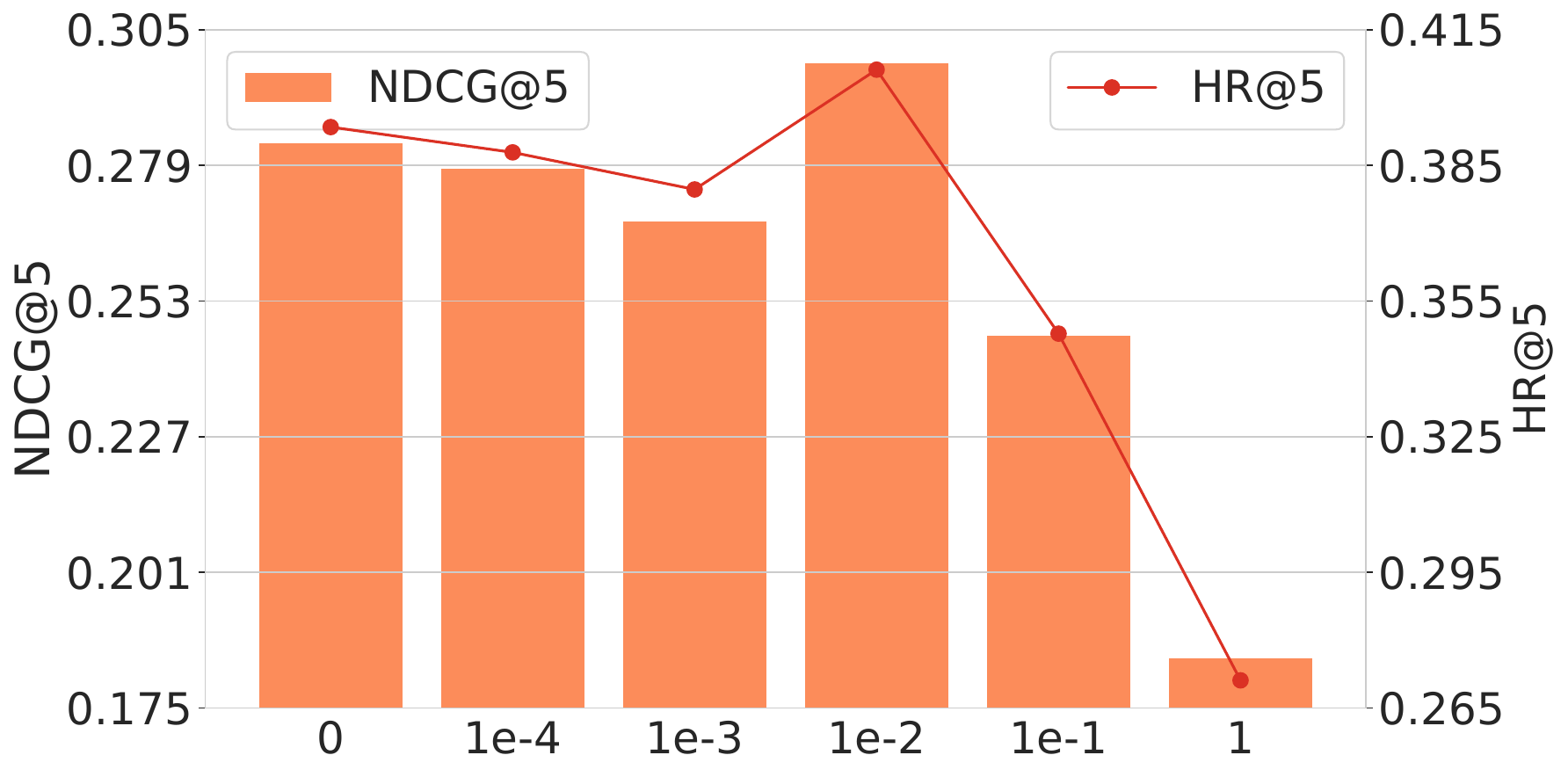}
     }
     \vspace{-8px}
     \caption{
     Impact of hyper-parameters $\lambda_{\mathrm{RQ\text{-}CL}}$, $\lambda_{\mathrm{U\text{-}CL}}$, and $\lambda_{\mathrm{His\text{-}CL}}$ on model performance, evaluated by NDCG@5 and HR@5.
     }
     \label{fig:hyper_param}
     \vspace{-0.5cm}
\end{figure*}

\subsubsection{Effectiveness of User-Code Graphs}
To enrich the representations of users with sparse search interactions, we construct user-code graphs using discrete codes (\S~\ref{sec:graph_construct}) and apply message passing (\S~\ref{sec:message_passing}) to propagate information from users with richer behaviors. To assess its effectiveness, we ablate the user-code graph, allowing the model to rely solely on users' \srcandrec histories for prediction. As shown by the ``w/o U-C Graph'' setting in Table~\ref{tab:ablation_result}, performance drops significantly, confirming the utility of leveraging rich user interactions to enhance sparse-user representations.

Furthermore, to facilitate more effective message passing, we introduce a user alignment loss $\mathcal{L}_{\mathrm{U\text{-}CL}}$ in Eq.~\eqref{eq:user_align} to align user embeddings and better capture cross-user similarity. Removing this loss (``w/o $\mathcal{L}_{\mathrm{U\text{-}CL}}$'') leads to notable degradation in performance, demonstrating the importance of embedding alignment in improving information propagation and similarity modeling.

\subsubsection{Effectiveness of Code and History Fusion}
In \S~\ref{sec:history_model}, we align and fuse the enhanced user \srcandrec code embeddings ($\mathbf{E}_{\tilde{s}}$ and $\mathbf{E}_{\tilde{r}}$) with the corresponding user \srcandrec histories. We separately evaluate the contributions of the alignment and fusion components.

For alignment, we introduce the contrastive loss $\mathcal{L}_{\mathrm{His\text{-}CL}}$ in Eq.~\eqref{eq:code_his_align} to align the code sequences with users' historical behaviors. As shown by the ``w/o $\mathcal{L}_{\mathrm{His\text{-}CL}}$'' setting in Table~\ref{tab:ablation_result}, removing this loss leads to noticeable performance degradation, underscoring the importance of aligning the two types of embeddings into a shared semantic space prior to fusion.

For the fusion step, as defined in Eq.~\eqref{eq:code_his_fusion}, we employ Multi-head Cross Attention (MCA) to integrate the code sequences with the historical behavior representations. As indicated by the ``w/o MCA'' setting in Table\ref{tab:ablation_result}, the absence of MCA results in reduced performance, validating the effectiveness of incorporating enhanced code embeddings to enrich the modeling of \srcandrec histories.

\subsection{Experimental Analysis}
We further conduct experimental analysis on the Qilin dataset to investigate the contributions of different modules.

\subsubsection{Ablation study across user groups with varying numbers of search interactions}
To further investigate the effectiveness of each module in addressing the search sparsity issue, we conduct ablation studies across user groups with different levels of search interaction sparsity, as shown in Figure~\ref{fig:ablation_group}.

We observe that removing the user-code graph module (``w/o U-C graph'' in Figure~\ref{fig:ablation_group}) causes a more substantial performance degradation for users with sparse search interactions. This indicates that message passing on the user-code graph can transfer useful information from users with rich search histories to those with sparse ones, thereby improving their representation quality.

Furthermore, removing the user alignment loss $\mathcal{L}_{\mathrm{U\text{-}CL}}$ (Eq.~\eqref{eq:user_align}) also leads to a more pronounced performance drop for sparse-search users. This highlights the critical role of the alignment loss in capturing user similarity, which enhances the effectiveness of message passing on the user-code graph. Moreover, this loss contributes to the learning of more discriminative user embeddings.

For the other modules, we generally observe that removing any of them leads to performance degradation across most user groups. This further validates the effectiveness and necessity of each component in the overall model.

\subsubsection{Embedding Visualization}
To gain deeper insights into the representations learned through message passing on the user-code graph (\S~\ref{sec:message_passing}), we visualize the user and code embeddings for both \srcandrec. Specifically, we analyze the user embeddings $\mathbf{E}_{\mathcal{U}}^s$ and $\mathbf{E}_{\mathcal{U}}^r$, along with the corresponding code embeddings $\mathbf{E}_{\widetilde{\mathcal{S}}}$ and $\mathbf{E}_{\widetilde{\mathcal{R}}}$. We employ t-SNE~\cite{van2008visualizing} to project the high-dimensional embeddings into a two-dimensional space, as shown in Figure~\ref{fig:code_emb}.

To assess the effectiveness of the user alignment loss, we compare the embedding distributions with and without the user contrastive loss $\mathcal{L}_{\mathrm{U\text{-}CL}}$ (Eq.~\eqref{eq:user_align}). Without this loss, the embeddings for \srcandrec are highly entangled, which can lead to redundant information being propagated through the graphs $\mathcal{G}_s$ and $\mathcal{G}_r$. In contrast, when the alignment loss is applied, the embeddings become more clearly separated, allowing the two graphs to model distinct user behavior patterns. This separation contributes to more effective message passing and ultimately leads to improved recommendation performance.

\subsubsection{Impact of Hyper-parameters}
We analyze the influence of the alignment loss weights $\lambda_{\mathrm{RQ\text{-}CL}}$, $\lambda_{\mathrm{U\text{-}CL}}$, and $\lambda_{\mathrm{His\text{-}CL}}$—corresponding to $\mathcal{L}_{\mathrm{RQ\text{-}CL}}$ (Eq.~\eqref{eq:rq_total_loss}), $\mathcal{L}_{\mathrm{U\text{-}CL}}$ (Eq.~\eqref{eq:total_loss}), and $\mathcal{L}_{\mathrm{His\text{-}CL}}$ (Eq.~\eqref{eq:total_loss})—on the final recommendation performance. Results are shown in Figure~\ref{fig:hyper_param}. During each analysis, the other two weights are fixed to their optimal values: $\lambda_{\mathrm{RQ\text{-}CL}} = 1\text{e-}4$, $\lambda_{\mathrm{U\text{-}CL}} = 1\text{e-}1$, and $\lambda_{\mathrm{His\text{-}CL}} = 1\text{e-}2$.

We find that a non-zero $\lambda_{\mathrm{RQ\text{-}CL}}$ consistently improves performance, highlighting the benefit of aligning user \srcandrec preference embeddings before quantization to better capture user similarity.
For $\lambda_{\mathrm{U\text{-}CL}}$, non-zero values occasionally degrade performance, suggesting that this loss requires careful tuning to effectively model user similarity in the user-code graph.
For $\lambda_{\mathrm{His\text{-}CL}}$, overly large values significantly harm performance, likely due to overemphasis on aligning code sequences with user histories at the expense of the main recommendation loss $\mathcal{L}_{\mathrm{rec}}$ (Eq.~\eqref{eq:total_loss}). Thus, this weight must be appropriately balanced to ensure optimal performance.

\section{Conclusion}
In this paper, we propose \ourname to address data sparsity in search-enhanced recommendation by leveraging users with rich search interactions to improve representations for users with sparse behaviors. We first use a LLM to summarize each user's \srcandrec preferences, which are then encoded and discretized via vector quantization. Users are connected to their codes, forming the user-code graphs where shared codes link similar users. Message passing on this graph enables knowledge transfer from rich to sparse users. We further introduce contrastive losses to enhance user similarity modeling. The refined user and code embeddings are finally integrated with user histories for prediction. Experiments on three real-world datasets show that \ourname consistently outperforms baselines, especially for users with sparse search activity.

\section*{GenAI Usage Disclosure}
During the writing of this paper, we used generative AI tools (e.g., ChatGPT) solely for the purpose of improving language clarity and grammar. No parts of the manuscript were generated directly by AI; all content, including ideas, experimental designs, results, and discussions, were conceived and written by the authors. The use of AI was limited to minor linguistic refinement and did not contribute to the creation of scientific content or analysis.

\bibliographystyle{ACM-Reference-Format}
\bibliography{ref}

\end{document}